\documentclass[twocolumn,tighten]{aastex63}
\usepackage{mathrsfs}
\usepackage[utf8]{inputenc}
\usepackage{mathtools}
\usepackage{amsmath}
\usepackage{amssymb}

\shorttitle{Sgr A* and Dark Matter Spike}
\shortauthors{Nampalliwar, S. et.al}

\graphicspath{{./}{figures/}}

\begin{document}

\title{Modelling the Sgr A* Black Hole Immersed in a Dark Matter Spike}

\author[0000-0002-9608-1102]{Sourabh Nampalliwar}
\affiliation{Theoretical Astrophysics,
Eberhard Karls Universität Tübingen, Tübingen, Germany}

\author[0000-0001-7156-4848]{Saurabh K.}
\affiliation{Department of Physics, Dyal Singh College, University of Delhi, India}

\author[0000-0003-0527-4177]{Kimet Jusufi}
\affiliation{Physics Department, State University of Tetovo, Ilinden Street nn, 1200, Tetovo, North Macedonia}

\author[0000-0002-3345-9905]{Qiang Wu}
\email{wuq@zjut.edu.cn}
\affiliation{Institute for Theoretical Physics and Cosmology, Zhejiang University of Technology,
Hangzhou, 310023 China}	

\author[0000-0001-9662-1546]{Mubasher Jamil}
\affiliation{Institute for Theoretical Physics and Cosmology, Zhejiang University of Technology,
Hangzhou, 310023 China}
\affiliation{School of Natural Sciences (SNS), National University of Sciences and Technology (NUST), Islamabad 44000, Pakistan}
 \affiliation{Canadian Quantum Research Center, 204-3002, 32 Ave, Vernon, BC, V1T 2L7, Canada}

\author[0000-0002-9015-9722]{Paolo Salucci}
\affiliation{SISSA/ISAS, International School for Advanced Studies, Via Bonomea 265, 34136, Trieste, Italy}

\begin{abstract}
In this paper, we investigate the effects of a dark matter (DM) spike on the neighborhood of Sgr A*, the black hole (BH) in the center of the Milky Way galaxy. Our main goal is to investigate whether current and future astronomical observations of Sgr A* could detect the presence of such a DM spike. At first, we construct the spacetime metric around a static and spherically symmetric BH with a DM spike, and later this solution is generalized for a rotating BH using the Newman-Janis-Azreg A\"{i}nou algorithm. For the static BH metric, we use the data of the S2 star orbiting the Sgr A* to determine and analyze the constraints on the two free parameters characterizing the density and the innermost boundary of the DM halo surrounding the BH. Furthermore, by making use of the available observational data for the DM spike density $\rho_\text{sp}$ and the DM spike radius $R_\text{sp}$ in the Milky Way galaxy, we consider a geometrically-thick accretion disk model around the Sgr A* BH  and demonstrate that the effect of DM distribution on the shadow radius and the image of the BH is considerably weak for realistic DM densities, becoming significant only when the DM density is of the order $\rho_\text{sp} \sim (10^{-19}-10^{-20})$ g/cm$^3$ near the BH. We further analyze the possibility of observing this effect with radio interferometry, simulating observations with an EHT--like array, and find that it is unlikely to be detectable in the near future.  
\end{abstract}



\section{Introduction}

Black holes (BHs) are some of the most fascinating astrophysical objects which perform manifestations of extremely strong gravity and high energy physics such as the formation of gigantic jets of particles, quasiperiodic oscillations, gravitational lensing and the disruption of nearby orbiting stars. For several decades, the direct evidence for the existence of BHs remained a mystery, however absence of evidence does not mean evidence of absence. In the past few decades, the observations of radiation from BH neighborhoods in X-rays~\citep{Fabian:1989ej,Laor:1991nc} and radio~\citep{Eckart:1996zz,Ghez:2000ay}, and more recently, the observations of the shadow of the event horizon of the M$87^{\star}$ supermassive BH by the EHT team \citep{eht} and the detection of the gravitational waves signal from mergers of two BHs by the LIGO/Virgo team \citep{ligo}, have provided convincing evidence for the presence of BHs in the universe. From a theoretical perspective, BHs serve as a lab to test various predictions of the theories of modified gravity, quantum gravity and other small or large distance corrections to the general relativity. 

In the astronomical survey conducted by the EHT team, the bright accretion disk surrounding the M$87^{\star}$ supermassive BH appears to be distorted due to the phenomenon of gravitational lensing. The region of accretion disk behind the BH also gets visible due to the bending of light by the BH. The shadow image helps in understanding the geometrical structure of the event horizon and the angular speed of the BH. The mass of a BH can be constrained, alternatively, by studying the dynamical processes such as orbital motion of nearby stars, however, measuring spin of a BH is more involved. In this regard, the iron line method, X-ray reflection spectroscopy and the continuum fitting methods are particularly suitable to constrain the spin of many astrophysical BHs~\citep{Reynolds:2013qqa,cb1,cb2}. 

 It is a well known fact that a DM halo envelopes every galaxy and even permeates the intergalactic medium. The distribution of DM near the region Sgr A$^{\star}$, in particular, is quite relevant to test and further constrain the predictions of general relativity (GR) and any modifications beyond GR, moreover, it will help in identifying the candidates of DM. If the central BH grows purely adiabatically by the standard accretion of dust and gases then the BH growth will create a sharp spike in the distribution of DM near the BH, whereas if the central BH is formed by the mergers of many small BHs then the spike would be relatively small while the amalgamation of the last two scenarios will result in an intermediate level spike, characterized by a power law~\citep{oleg}. Gondolo and Silk proposed that if the galactic center contains DM then the presence of a supermassive BH in the galactic center would create a cusp in the distribution of the DM which they termed as a DM spike, with the DM density $\rho\sim r^{-\gamma},$ with $0<\gamma<2$~\citep{silkg}. The work of Gondolo \& Silk does 
not take into account general relativistic effects of DM spike close to the black hole. These effects 
were studied in \citep{Sadeghian:2013laa,Ferrer:2017xwm} for Schwarzschild and Kerr black hole. A noticeable feature of the DM spike would be the annihilation of the particle DM in the high energy regime, thereby forming a spike luminosity depending on the density of the inner halo \cite{Merritt:2002vj,Fields:2014pia}. Constraints on the spiky DM model in different galactic systems including M87 and Milky Way, have been dealt in Refs.~\citep{mark,silk2,lac,kuh}. Spinning and non-spinning BHs and their shadows in the presence of DM have been extensively studied before~\citep{jamil1,jamil2, jamil3, jamil4,Hou:2018bar,Boshkayev:2020kle,Konoplya:2019sns}.
 
In the case of the Milky Way galaxy, the corresponding DM density profile for the inner regions $r<1$~kpc (kilo parsec) is very uncertain. One must rely on extrapolations to the galactic center by the astronomical observations and/or employ state-of-the-art N-body $+$ hydrodynamics simulations obtained at the kpc scale. At $r\ll 100$~pc from the galactic center, the BH severely modifies the distribution of DM profile. In an environment with a very high DM density, the adiabatic growth of the BH due to accretion of particles at $100$~pc scales induces  a steep increase in the DM density, leading to a spike. The density can reach up to values $6 \times 10^{-10}$ g/cm$^{3}$, which is much bigger then the largest density measured in Milky Way from its internal kinematics. In principle, a steepening of the density as $r^{-2.4}$ could continue downwards to a radius of order several times the Sgr A$^\star$ Schwarzschild radius $R_S$ \citep{Nishikawa:2017chy,Kavanagh:2020cfn}.
 
In this work, we consider a model based on the idea that the central regions, including the central BH, is surrounded by a DM spike satisfying a power law density profile. 
In particular, it is interesting to note that the stellar components dominate the gravitational potential in the region between $\sim$ 100pc to $\sim$ kpc. Beyond $r \sim$ kpc, the usual NFW-like "halo" or the Burkert--Salucci profile takes over. Then within $r \ll  100$pc, the dark matter spikes dominate the relevant dynamics.
Our aim in the present work is to explore the effects of the DM spike on the orbit of the S2 star, on the shadow radius of the Sgr~A$^\star$ BH, and the effect of DM spike on the \textit{images} of the central region of Sgr~A$^\star$ surrounded by a geometrically-thick accretion and radiative disk model. Very importantly, we aim to simulate the images of Sgr~A$^\star$ BH with the DM effect using an array of radio interferometric array. 

The plan of the paper is as follows: In Sec.~\ref{sec:profile}, we use the DM spike profile to compute the metric around a static BH.  In Sec.~\ref{sec:s2}, we discuss the constraints on the density and the inner edge of DM distribution in the Milky Way galactic center using data from the orbit of S2. In Sec.~\ref{sec:shadows}, we generalize the metric of Sec.~\ref{sec:profile} to rotating BHs and study the shadows around these BHs for different DM profiles and strengths. The effects of DM on the BH images are analyzed in Sec.~\ref{sec:images}, using a geometrically-thick and optically-thin accretion disk model. In Sec.~\ref{sec:radio}, we explore the detectability of these DM effects with simulations of radio interferometric observations of the Sgr~A$^\star$ BH. We conclude in Sec.~\ref{sec:conclude}. Note that we shall use the natural units $G=c=1$ throughout the paper.

\section{\label{sec:profile}The dark matter spike profile}
We start by considering a BH, whose mass we denote by $M_\text{BH}$, residing in the center of a DM halo which initially has a power law density profile near the Galactic center given by~\citep{Nishikawa:2017chy} 
\begin{equation}
\rho_\text{DM}(r) \simeq \rho_0 \left(\frac{r_0}{r}\right)^{\gamma},
\label{rhodm}
\end{equation}
where $\gamma$ is the power-law index and $\rho_0$ and $r_0$  are halo parameters. As shown in Refs.~\citep{Nishikawa:2017chy,Kavanagh:2020cfn}, this will lead to the formation of a DM spike of radius 
\begin{equation}
	R_\text{sp}(\gamma, M_\text{BH}) = \alpha_\gamma r_0 \left( \frac{M_\text{BH}}{\rho_0 r_0^3}\right)^{1/(3-\gamma)},
\end{equation}
where the normalization $\alpha_{\gamma}$ is numerically calculated for each power-law index $\gamma$. We assume that the DM spike was formed as a result of adiabatic growth of BH which enhances the central density of the host halo. The DM distribution in the spike region is given by~\citep{Kavanagh:2020cfn} 
\begin{equation}
	\rho^\text{sp}_\text{DM}(r) =\rho_\text{sp}\, {\left(\frac{R_\text{sp}}{r}\right)}^{\gamma_\text{sp}},
    \label{eq:rhosp}
\end{equation}
where 
$\gamma_\text{sp} = (9-2\gamma)/(4-\gamma)$, and $R_s =2  M_\text{BH} \simeq 2.95 \ (M_\text{BH}/\, { M}_\odot)\,{\rm km}$ is the Schwarzschild radius of the BH.
Note that this DM density profile differs from the Navarro-Frenk-White (NFW) density profile, which is motivated by the numerical simulations of the collisionless DM particles in the galactic halos, for $\gamma=1$ and 
$\gamma=0$ and $M_\text{BH}=10^5\, M_{\odot}$ or $M_\text{BH}=10^6\, M_{\odot}$. 
It is interesting that the DM density is enhanced by several orders of magnitude in the spike region, and it is therefore worth investigating whether this could have a significant impact on the observable signatures from the central BH in the Milky Way. The first step towards this investigation is to construct a metric background on which particles, both massive and massless, will travel.


Using the density profile given by Eq.~(\ref{rhodm}), we obtain
\begin{eqnarray}\label{Msp}
M^\text{sp}_\text{DM}(r)=4 \pi \int_{r_\text{b}}^{r} r'^2 \rho_\text{DM}^\text{sp}(r') dr',\,\,\,\,r_\text{b} \leq r' \leq R_\text{sp}
\end{eqnarray}
and the total mass is
\begin{eqnarray}\label{Mr}
M(r)=M_\text{BH}(r)+M^\text{sp}_\text{DM}(r).
\end{eqnarray}
Hence for the DM spike we find
\begin{eqnarray}\label{Mspdm}
M^\text{sp}_\text{DM}(r)=\frac{4 \pi \rho_\text{sp} r_\text{b}^3}{\gamma_\text{sp}-3} \left(\frac{R_\text{sp}}{r_\text{b}}\right)^{\gamma_\text{sp}}-\frac{4 \pi \rho_\text{sp} r^3}{\gamma_\text{sp}-3} \left(\frac{R_\text{sp}}{r}\right)^{\gamma_\text{sp}}.
\end{eqnarray}
We assume that the DM is restricted in the domain $r_\text{b} \leq r \leq R_\text{sp} $, where  $r_\text{b}$ is the inner edge of the DM spike and $M_\text{BH}$ is the mass of BH.
We therefore have:  
\begin{eqnarray} 
M(r)=\begin{cases}
M_\text{BH}, & r\leq r_{b};\\
M_\text{BH}+ M_\text{DM}^\text{sp}(r), & r_{b}\leq r\leq R_\text{sp};\\
M_\text{BH}+ M_\text{DM}, & r>R_\text{sp}
\end{cases}
\end{eqnarray}

 We notice that the DM mass shell is essentially a fixed mass which depends on the density and its location. In fact, at large scales, $r\gg R_\text{sp}$, the spacetime can be matched with either the Burkert-Salucci profile~\citep{saluci1,Burkert}
\begin{eqnarray}\label{BS}
\rho(r)=\frac{\rho_0 r^3_0}{(r+r_0)(r^2+r_0^2)},
\end{eqnarray}
 or the NFW profile \citep{nfw}
 \begin{eqnarray}\label{NFW}
 \rho_\text{NFW}(r)=\rho_s \frac{r_s}{r\left(1+\frac{r}{r_s}\right)^2}.
 \end{eqnarray}
 Obviously, the effect of DM, on the shadow for instance, depends on the mass distribution very close to the BH in the galactic center. Hence, we are going to neglect the DM which is far away from the spike, i.e., in the region  $r\gg R_\text{sp}$. Furthermore, we set $\gamma_\text{sp}$ in the range $\gamma_\text{sp} \in [0,3]$, since at $\gamma_\text{sp}=3$ there is an apparent singularity. Typically, $\gamma_\text{sp}>2$ is never seen in the centers of galaxies.

Now, we proceed to solve the Tolman-Oppenheimer-Volkoff (TOV) equation in the DM halo with a BH. We begin with a spherically symmetric system: the BH is assumed to be Schwarzschild, while the DM profile is spherically symmetric by construction. (Generalization to a rotating BH is presented below, in Sec.~\ref{sec:shadows}.) To analyze the properties of the system composed of a BH and DM envelope, we start with the generic line element in the standard static and spherically symmetric form as follows
\begin{equation}\label{eq:le}
d s^2=-e^{N(r)} d t^2 + e^{\Lambda(r)}d r^2 + r^2 \left(d \theta^2 + \sin^2 \theta d\phi^2\right),
\end{equation}
where $(t,r,\theta,\phi)$ are the usual temporal and spatial coordinates, and $N(r) $ and $\Lambda(r)$ are the sought metric functions. Within the region $r_{b}\leq r\leq R_{sp}$, we can always choose $e^{\Lambda(r)}=g(r)^{-1}$, where 
\begin{eqnarray}\label{eq:gfunc}\notag
g(r)&=& 1-\frac{2 M(r)}{r}=1-\frac{2 M_\text{BH}}{r}-\frac{8 \pi \rho_\text{sp} r_\text{b}^3}{r(\gamma_\text{sp}-3)} \left(\frac{R_\text{sp}}{r_\text{b}}\right)^{\gamma_\text{sp}}\\
&+&\frac{8 \pi \rho_\text{sp} r^2}{\gamma_\text{sp}-3} \left(\frac{R_\text{sp}}{r}\right)^{\gamma_\text{sp}}.
\end{eqnarray}

In order to get a stable (static) spherical layer (envelope) of DM around the BH one has to match the inner BH spacetime with the outer, matter filled, solution describing DM at the boundary $r_\text{b}$. This can be done following the same procedure to find interior solutions for the Schwarzschild spacetime, since, as it is well known, the matching conditions to be satisfied are the same
\begin{eqnarray}\label{bc}
e^{N(r_\text{b})}=1-\frac{2M_\text{BH}}{r_\text{b}}.
\end{eqnarray}
for $r\leq r_\text{b}$. In particular we need to solve the TOV equations, which read as follows
\begin{eqnarray}\label{tov}
\frac{d P(r)}{d r}&=&-(\rho(r) +P(r)) \frac{M(r)+4 \pi r^3 P(r)}{r(r-2 M(r))} ,\\
\frac{d N(r)}{dr}&=& 2\,\frac{M(r)+4 \pi r^3 P(r)}{r(r-2 M(r))} ,
\end{eqnarray}
where $P(r)$ is the DM pseudo-pressure that can be defined also for collisionless particles \citep{scot}.  In particular, we shall consider the equation of state for the DM spike given by 
\begin{equation}
P(r)=\omega \rho_\text{sp}(r).
\end{equation} 
In what follows we are going to consider two cases: $\omega=0$ and $\omega=1$, respectively.
\subsection{Case I: $\omega=0$}
Using this equation of state which is relevant for modelling cold DM spike and introducing $f(r)=e^{N(r)}$, we solve the TOV equation for the metric function  by approximating the integral in leading order of $\rho_\text{sp}$ and fixing the constant of integration using the matching condition in Eq.~(\ref{bc}). We find
\begin{equation}\label{eq:ffunc}
  f(r)= 1-\frac{2M_\text{BH}}{r}-\exp \left[ - \frac{8 \pi \rho_\text{sp} r_\text{b}^2  \left(\frac{R_\text{sp}}{r_\text{b}}\right)^{\gamma_\text{sp} } }{(\gamma_\text{sp}-2)}   \right]
\end{equation}
\begin{equation}\notag
   + \exp{\left[-\frac{8 \pi \rho_\text{sp} \left( r_\text{b}^3 (\gamma_\text{sp}-2) \left(\frac{R_{sp}}{r_\text{b}}\right)^{\gamma_\text{sp} } -r^3  \left(\frac{R_\text{sp}}{r}\right)^{\gamma_\text{sp}}\right)}{r\,(\gamma_\text{sp}-3)(\gamma_\text{sp}-2)}\right]},
\end{equation}
which is valid for $r_{b}\leq r\leq R_{sp}$. It can be verified that at the inner boundary of the spike,
\begin{equation}
   \lim_{r \to r_\text{b}} f(r)= 1-\frac{2 M_{BH}}{r_\text{b}}.
\end{equation}
We can also approximate Eq.~(\ref{eq:ffunc}) in leading order of $\rho_\text{sp}$ as follows
\begin{eqnarray}\label{frsim}
f(r)& \simeq & 1-\frac{2 M_\text{BH}}{r}+\frac{8 \pi \rho_\text{sp} r^2}{(\gamma_\text{sp}-3)(\gamma_\text{sp}-2)} \left(\frac{R_\text{sp}}{r}\right)^{\gamma_\text{sp}}\\\notag
&+&\frac{8 \pi \rho_\text{sp} r_\text{b}^2 \Big[r\,(\gamma_\text{sp}-3)-r_\text{b}(\gamma_\text{sp}-2)\Big]}{r\,(\gamma_\text{sp}-3)(\gamma_\text{sp}-2)}  \left(\frac{R_\text{sp}}{r_\text{b}}\right)^{\gamma_\text{sp}}.
\end{eqnarray}
It is interesting to see that there are two apparent singularities for $\gamma_\text{sp}=3$ and $\gamma_\text{sp}=2$. As an aside, it is relevant to consider the limit of metric function (\ref{eq:ffunc}) for large but finite values of $r\gg r_\text{b}$, i.e., an observer located at $r \sim R_\text{sp}$, and then approximating in leading order terms on $\rho_\text{sp}$, yielding
\begin{equation}\label{c1}
f(r) \simeq 1+ 32 \pi \rho_\text{sp}\Big[  r_\text{b}^2 \left(\frac{R_\text{sp}}{r_\text{b}}\right)^{9/4}+\frac{r_\text{b}^3\left(\frac{R_\text{sp}}{r_\text{b}}\right)^{9/4}-4R_\text{sp}^3 }{3R_\text{sp}}\Big],
\end{equation}
for the case $\gamma_\text{sp}=9/4$ (corresponding to $\gamma = 0$), and 
\begin{equation}\label{c2}
f(r) \simeq 1+ 24 \pi \rho_\text{sp}\Big[  r_\text{b}^2 \left(\frac{R_\text{sp}}{r_\text{b}}\right)^{7/3}+\frac{r_\text{b}^3\left(\frac{R_\text{sp}}{r_\text{b}}\right)^{7/3}-3R_\text{sp}^3 }{2R_\text{sp}}\Big],
\end{equation}
for the case $\gamma_\text{sp}=7/3$ (corresponding to $\gamma = 1$). In the above approximation, we have also neglected the effect of BH mass. We can see that the DM spike implies that our spacetime is not asymptotically flat. 
\subsection{Case II: $\omega=1$}

The case $\omega=1$ is called the stiff (or causal) equation of state in the literature. It is commonly used to study ultra-relativistic regimes which might occur in the early universe or in the core of ultra-compact objects such as neutron stars. It
may also arise in certain cosmological models where dark matter is made from relativistic self-gravitating
Bose-Einstein condensates \citep{Chavanis:2014lra}. In the present context, it might be relevant for the dark matter spike which lies very close to the central BH and hence its equation of state is not known precisely. It is an educated guess that the SBH at MW galaxy center might push the spike DM density closer to ultra-relativistic limit which is approximated as $\omega=1$ \citep{Harko:2014zma}.

Using this particular equation of state, the solution from the TOV equation for the metric function can be approximated in leading order of $\rho_\text{sp}$ as follows
 \begin{equation}\label{omega=1}
  f(r)= 1-\frac{2M_\text{BH}}{r}-\exp \left[ - \frac{16 \pi \rho_\text{sp} r_\text{b}^2  \left(\frac{R_\text{sp}}{r_\text{b}}\right)^{\gamma_\text{sp} } }{(\gamma_\text{sp}-2)}   \right]+\notag
\end{equation}
\begin{equation}
   \exp{\left[-\frac{8 \pi \rho_\text{sp} \Big( r_\text{b}^3 (\gamma_\text{sp}-2) (\frac{R_\text{sp}}{r_\text{b}})^{\gamma_\text{sp} } +r^3 (\gamma_\text{sp}-4)  (\frac{R_\text{sp}}{r})^{\gamma_\text{sp}}\Big)}{r\,(\gamma_\text{sp}-3)(\gamma_\text{sp}-2)}\right]}
\end{equation}
which is valid for $r_{b}\leq r\leq R_\text{sp}$. It can be further approximated as
\begin{eqnarray}
f(r)& \simeq & 1-\frac{2 M_\text{BH}}{r}-\frac{8 \pi (\gamma_\text{sp}-4)\rho_\text{sp} r^2}{(\gamma_\text{sp}-3)(\gamma_\text{sp}-2)} \left(\frac{R_\text{sp}}{r}\right)^{\gamma_\text{sp}} \\\notag
&+&\frac{8 \pi \rho_\text{sp} r_\text{b}^2 \Big[2\,r\,(\gamma_\text{sp}-3)-r_\text{b}(\gamma_\text{sp}-2)\Big]}{r\,(\gamma_\text{sp}-3)(\gamma_\text{sp}-2)} \left(\frac{R_\text{sp}}{r_\text{b}}\right)^{\gamma_\text{sp}}
\end{eqnarray}
Again, we see that at the limit $r=r_\text{b}$, our metric reduces to the condition in Eq.~(\ref{bc}). We see that there is slight difference between the metric functions in the $\omega=0$ and the $\omega=1$ case, respectively, therefore it will be worthwhile to explore and see the effect on the observables. Taking the limit $r \sim R_\text{sp}$ and then approximating in leading order terms on $\rho_\text{sp}$, we find
\begin{equation}
f(r) \simeq 1+ 64 \pi \rho_\text{sp}\Big[  r_\text{b}^2 \left(\frac{R_\text{sp}}{r_\text{b}}\right)^{9/4}+\frac{r_\text{b}^3\left(\frac{R_\text{sp}}{r_\text{b}}\right)^{9/4}-7R_\text{sp}^3 }{6R_\text{sp}}\Big],
\end{equation}
for $\gamma_\text{sp}=9/4$, and 
\begin{equation}
f(r) \simeq 1+ 48 \pi \rho_\text{sp}\Big[  r_\text{b}^2 \left(\frac{R_\text{sp}}{r_\text{b}}\right)^{7/3}+\frac{r_\text{b}^3\left(\frac{R_\text{sp}}{r_\text{b}}\right)^{7/3}-5R_\text{sp}^3 }{4R_\text{sp}}\Big],
\end{equation}
for $\gamma_\text{sp}=7/3$. We can see that, in this case, the effect is slightly different compared to Eqs.~(\ref{c1}) and (\ref{c2}). As a final note, we point out that at large distances outside the DM spike radius, i.e., $r\gg R_\text{sp}$, one should match the DM spike metric with the outside metric obtained via the NFW profile or Burkert-Salucci profile. However, simply by means of the continuity of the metric, we still expect the contribution of the nontrivial topology to have an effect outside $R_\text{sp}$. This situation is similar to the spacetime outside the global monopole which can be described by the global conical topology. Compared to the global monopole metric, one can check that there is a crucial sign difference here which results from the DM pressure or simply by means of the TOV equation used in our setup.

Having derived a metric background that describes a spherically symmetric BH in a DM spike, we look at our first observable: the orbit of a star around the BH.
\section{\label{sec:s2}Constraints on parameters from the star S2 orbit about the Sgr A$^{\star}$ BH}
\begin{figure}
\includegraphics[scale=0.82]{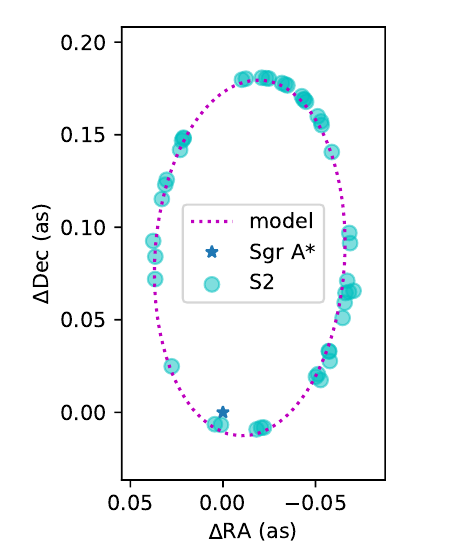}
\centering
    \caption{Simulated and observed orbit of S2 around Sgr A$^\star$. We use the observational data in \citep{Do:2019}}
    \label{orbit}
\end{figure}
It is well-known that several stars, in particular the S cluster orbits the Sgr A$^{\star}$ central BH \citep{nucita, nucita:1, nucita:2}, can be utilized to investigate the physical characteristics of the Sgr A$^{\star}$ BH, such as its proper motion, mass and angular momentum.  Recently it has been reported in \citep{Fragione:2020khu} that Sgr A* BH in the center of Milky Way is slowly spinning. They have estimated that the dimensionless spin vector has the upper bound $|\chi|<0.1$ (as per their assumptions $|\chi|=0$ and $|\chi|=1$ correspond to non-spinning and maximally spinning BHs respectively). This estimate has been made by measuring the effects of spin precession and frame dragging experienced by the S-cluster stars in the MW center.  There have been reportedly two known methods to constrain the parameters of Sgr A$^{\star}$. The first method pertains analyzing the periastron shift of the S2 star orbiting nearest the central BH among the stars of S cluster while the second method involves the study of retro-lensing of the brightest and innermost stars orbiting Sgr A$^{\star}$ \citep{nucita1, nucita2}. The same observations can also help constraining the DM distribution near the galactic center \citep{alex1, alex2}. The motion of S cluster stars around Sgr A$^{\star}$ have set strong constraints on the mass of the compact object at the centre of the Milky Way galaxy, which is assumed to be a massive BH with a mass of about $4.1\times 10^6 M_{\odot}$ \citep{Gillessen:2017}. However, some latest observations of the motion of G2 and S2 stars have discredited the nature of Sgr A$^{\star}$ to be a BH, therefore some authors have modelled the central gravitational object as a dense core constituted of a diluted halo of fermionic DM only \citep{rueda}.  Although the nature of the inner region (sub-pc), especially inside the orbits of stars of S-cluster still remains unclear, it is hoped that some S-stars, for instance, S2 could give the constraints about extended mass distribution of the DM profile. In this section, we use the S2 orbit data, collected over the last few decades \citep{Do:2019}, to fit the DM profile in Eq. (\ref{eq:rhosp}). We parameterize the model using two parameters, inner edge radius of the DM halo $r_\text{b}$ and the characteristic density $\rho_\text{b}$ given  by
\begin{equation}
	\rho_\text{b} =\rho_\text{sp}\, {\left(\frac{R_\text{sp}}{r_\text{b}}\right)}^{\gamma_\text{sp}},
    \label{eq:rhob}
\end{equation}
To obtain the S2 orbit, we solve the equations of motion Eq.(C.1) in \citep{rueda} numerically with the metric coefficients from Eqs.~(\ref{eq:gfunc}) and (\ref{frsim}) by using the python integrator LSODA \citep{lsoda} which apply the Adams/BDF method with automatic stiffness detection and switching. The orbit is projected to sky plane using the transformation Eq.(C.9) in \citep{rueda}. Fitting the model with observational data, we fixed the orbit parameters such as inclination angle ($i$), argument of periapsis ($\omega'$), angle to the ascending node ($\Omega$) using the values in \citep{Do:2019}, then varied the semi-major axis ($a$) and eccentricity ($e$) of the orbit and two free model parameters $r_\text{b}$ and $\rho_\text{b}$. The best-fitting values and confidence level are derived from the  Monte-Carlo-Markov Chains analysis using the open-source python package \textit{emcee} \citep{emcee} which apply the affine invariant ensemble methods to speed up the sampling.

\begin{figure*}
\centering
\includegraphics[width=3.4in]{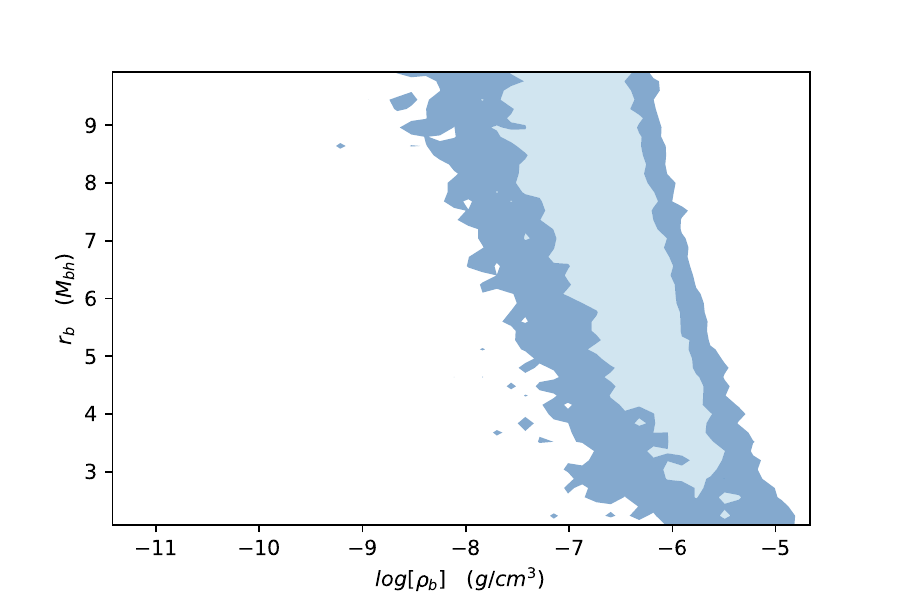}
\label{a}
\includegraphics[width=3.4in]{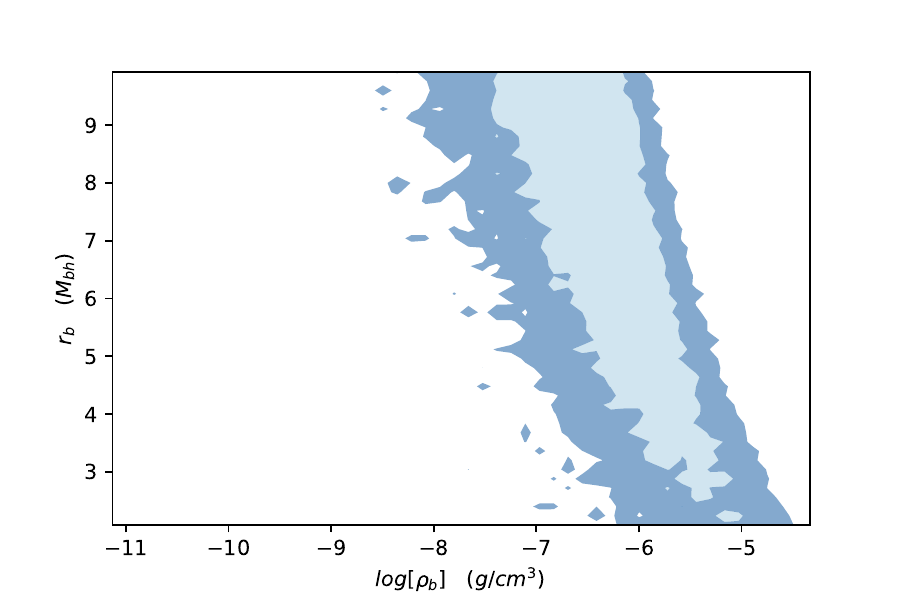}
    \caption{The constrained DM profile parameters $\rho_\text{b}$ and $r_\text{b}$ with 68\% and 96\% confidence contours for $\gamma_\text{sp}=9/4$ (left) and $\gamma_\text{sp}=7/3$ (right).}
    \label{contour}
\end{figure*}

The observational data and the best-fitting orbit for the S2 are shown in Fig.~\ref{orbit}, where the star denotes the position of Sgr A$^\star$. The resulting parameter constraints for $\rho_\text{b}$ (in base-10 log scale) and $r_\text{b}$ are shown in Fig.~\ref{contour} for the Case I ($\omega = 0$) are also similar for Case II ($\omega = 1$). We took the uniform priors for $r_\text{b} \sim [2.0,10.0]$ $M_{BH}$ and $\rho_{b}\sim[3.0\times 10^{-12},2.5\times10^{-5}]$ g/cm$^3$.  Fig.~\ref{contour}, shows a clear degeneracy between the parameters $\rho_b$ and $r_b$ due to the fact that the astronomical data fit the orbital shape of S2 star quite well. Moreover, the data lacks the ability to constrain the modified gravity or beyond-GR effects as given in our model. We hope that future observations of stellar orbital precession will offer the opportunity to better constrain modified gravity theories.

For the values of $\gamma_\text{sp} =  9/4$ and $7/3$ respectively, the best fitting values and $96\%$ confidence level of parameters $r_\text{b}$ and $\rho_\text{b}$ in Case I are presented in Table~\ref{values}.

Taking the case $\gamma_\text{sp}=9/4$ as an example, we find that the inner edge $r_\text{b}$ lies within the spatial size of few $M_{BH}$, which covers the region between S-stars and the Schwarzschild radius of BH. The density at $r_\text{b}=9.4 M_{BH}$ is of the order $\rho_\text{b}\sim 4.5\times 10^{-5}$ g/cm$^3$ at the $2 \sigma$ level, which means that the DM density is enhanced by several orders of magnitude in the spike region close to the BH. However, from the results of parameter constraints shown in Table~\ref{values}, we see that the DM is located outside the photon sphere, although there is some uncertainty in constraining $r_\text{b}$ using the motion of the S2 star. In what follows, we will generalize the metric background to include rotation and explore some more observables.

\begin{deluxetable}{ccc}
\tablecaption{Best fitting values and the 96\% confidence level of model parameters for  Case I using S2 star orbital data. We set the BH mass  $M_\text{BH}=4.1\times 10^6 M_\odot$}
\tablehead{\colhead{{$\gamma_\text{sp}$}} & \colhead{{$\rho_{b}$ $[g/cm^3$ $\times 10^{-7}$] }} & \colhead{{$r_\text{b}$ $[M_{BH}$]}}} 
\startdata
$9/4$ & $2.5^{+41.5}_{-1.7}$  & $9.4^{+0.5}_{-6.9} $   \\
$7/3$ & $5.3^{+88.7}_{-1.6}$  & $9.4^{+0.5}_{-7.0} $   \\
\enddata

\end{deluxetable}
\label{values}

 \section{\label{sec:shadows}Shadow of a rotating BH in a dark matter spike}
We start by deriving the effective metric of a rotating BH in a DM spike by using a general seed static and spherically symmetric metric:
	\begin{equation}\label{1}
	ds^2=-f(r)dt^2+\frac{dr^2}{g(r)}+h(r) (d\theta^2+\sin^2\theta d\phi^2).
	\end{equation}
	We can use the Newman-Janis algorithm by transforming the metric from Boyer-Lindqiust coordinates $(t, r, \theta, \phi)$ to Eddington-Finkelstein coordinates $(u, r, \theta, \phi)$ and apply the Newman-Janis algorithm by complexifying the radial coordinate $r$. In particular, we shall follow the Newman-Janis algorithm modified by Azreg-A\"{i}nou's non-complexification procedure \citep{Azreg-Ainou:2014pra},  by which one drops the complexification step of the Newman-Janis algorithm first by using the null tetrads 
	\begin{eqnarray}
	l^{a}&=&\delta^{a}_{r},\nonumber \\
	\label{e} n^a&=&\sqrt\frac{G}{F}\delta^{a}_{u}-\frac{G}{2}\delta^{a}_{r},\\\
	m^{a}&=&\frac{1}{\sqrt{2H}} \left[(\delta^{a}_{u}-\delta^{a}_{r})\dot{\iota}{a}\sin\theta+\delta^{a}_{\theta}+\frac{\dot{\iota}}{\sin\theta}\delta^{a}_{\phi}\right],\nonumber
	\end{eqnarray}
	where the functions $f(r)$, $g(r)$ and $h(r)$ transform to new functions $F=F(r,a,\theta)$, $G=G(r,a,\theta)$ and $H=H(r,a,\theta)$, respectively. We like to emphasize that the functions $f(r)$ and $g(r)$ are adopted from Section II for later analysis. Without going into details one can show  \citep{Azreg-Ainou:2014pra}
	\begin{eqnarray}
	\label{FG}
	F &\equiv&\frac{\left(g(r)h(r)+a^2\cos^2\theta\right)H}{\left(k(r)+a^2\cos^2\theta\right)^2},\\
	G &\equiv&\frac{\left(g(r)h(r)+a^2\cos^2\theta\right)}{H},
	\end{eqnarray}
	and the effective rotating BH metric in Kerr-like coordinates turns out to be 
	\begin{eqnarray}\label{eq:rotmet}\notag
	ds^2 &=& \frac{H}{\rho^2} \Big[-\frac{\Delta}{\rho^2}(dt-a\sin^2\theta d\phi)^2+\frac{\rho^2}{\Delta}dr^2+\rho^2d\theta^2 \\
	&+&\frac{\sin^2\theta}{\rho^2}(a dt-(k+a^2)d\phi)^2 \Big],  
	\end{eqnarray}
	with
		\begin{eqnarray}
	\Delta(r)&=& g(r)h(r)+a^2,\\
	k(r)&=&\sqrt{g(r)/f(r)}\,h(r)\\
		\rho^2(r)&=&k+a^2\cos^2\theta,
	\end{eqnarray}
with $a=L_{BH}/M_{BH}$ being the specific angular momentum (rotation parameter) and $M_{BH},\,L_{BH}$ being the mass and angular momentum of the spinning BH, respectively. 
The $H(r, \theta, a)$ function is still arbitrary and can be chosen so that the cross-term of the Einstein tensor $G_{r \theta}$, for a physically acceptable rotating solution,
identically vanishes, i.e., $G_{r \theta}=0$. The latter constraint yields the differential equation \citep{Azreg-Ainou:2014pra}
\begin{eqnarray}
(k(r)+a^2 y^2)^2(3 H_{,r}H_{,y^2}-2H H_{,r y^2})=3 a^2 k_{,r}H^2,
\end{eqnarray}
where $y\equiv\cos \theta$. Note that $H(r)$ is an unknown function. However, by setting $h(r)=r^2$ which is valid for most static and spherically symmetric BH spacetimes, the above equation has the following solution: 
\begin{eqnarray}\label{eq:hfunc}
H\equiv \sqrt{g(r)/f(r)} \, r^2+a^2\cos^2\theta.
\end{eqnarray}


Using this rotating metric, we explore the next observable: the BH \textit{shadow} (more precisely, the apparent boundary~\citep{Bardeen:1973tla,Luminet:1979nyg}). To this end, we solve the equations of motion of a photon in this background. We begin with the Hamilton-Jacobi equations, which can be written as
\begin{equation}
\frac{\partial \mathcal{S}}{\partial \sigma}=-\frac{1}{2}g^{\mu\nu}\frac{\partial \mathcal{S}}{\partial x^\mu}\frac{\partial \mathcal{S}}{\partial x^\nu},
\label{eq:HJE}
\end{equation}
where $\sigma$ is an affine parameter and $\mathcal{S}$ denotes the Jacobi action. There are two conserved quantities: the energy $E=-p_t$, and the conserved angular momentum $J=p_\phi$ (about the axis of symmetry). In order to find a separable solution of Eq.~(\ref{eq:HJE}), we can express the action in terms of the known constants of motion as follows
\begin{equation}
\mathcal{S}=\frac{1}{2}m ^2 \sigma - E t + J \phi + \mathcal{S}_{r}(r)+\mathcal{S}_{\theta}(\theta),
\label{eq:action_ansatz}
\end{equation}
where $m$ denotes the mass of the test particle. However, for a photon we take $m=0$. Putting Eq.~(\ref{eq:action_ansatz}) in Eq.~(\ref{eq:HJE}), it is straightforward to derive the following equations of motion \citep{Shaikh:2019fpu}
\begin{equation}
H\frac{dr}{d\sigma}=\pm \sqrt{R(r)},
\label{eq:r_eqn}
\end{equation}
\begin{equation}
H\frac{d\theta}{d\sigma}=\pm \sqrt{\Theta(\theta)},
\label{eq:theta_eqn}
\end{equation}
where
\begin{gather}
R(r)=\left[\mathcal{X}(r)E-aJ\right]^2-\Delta(r)\left[\mathcal{K}+\left(J-aE\right)^2\right],\\
\Theta(\theta)=\mathcal{K}+a^2E^2\cos^2\theta-J^2\cot^2\theta,
\end{gather}
where $\mathcal{X}(r)=(\sqrt{g(r)/f(r)}\, r^2+a^2)$, and $\Delta(r)$ is defined by Eq. (29), while $\mathcal{K}$ is known as the Carter separation constant, which is another constant of motion. If we define $\xi=J/E$ and $\eta=\mathcal{K}/E^2$, one can show that the unstable circular photon orbits in the general rotating spacetime must satisfy $R(r_\text{ph})=0$, $R'(r_\text{ph})=0$ and $R''\geq 0$, where $r=r_\text{ph}$ represents the radius of the unstable photon orbit. Using the above conditions one can show \citep{Shaikh:2019fpu}
\begin{gather}
\left[\mathcal{X}(r_\text{ph})-a\xi\right]^2-\Delta(r_\text{ph})\left[\eta+\left(\xi-a\right)^2\right]=0,\label{eq:Req0}\\
2\mathcal{X}'(r_\text{ph})\left[\mathcal{X}(r_\text{ph})-a\xi\right]-\Delta'(r_\text{ph})\left[\eta+\left(\xi-a\right)^2\right]=0.
\label{eq:Rpeq0}
\end{gather}

After some algebraic manipulations, we can eliminate $\eta$ from the last two equations and solve for $\xi$, to obtain \citep{Shaikh:2019fpu}
\begin{gather}
\xi=\frac{\mathcal{X}_\text{ph}\Delta'_\text{ph}-2\Delta_\text{ph}\mathcal{X}'_\text{ph}}{a\Delta'_\text{ph}},\label{eq:xi}\\
\eta=\frac{4a^2\mathcal{X}'^2_\text{ph}\Delta_\text{ph}-\left[\left(\mathcal{X}_\text{ph}-a^2\right)\Delta'_\text{ph}-2\mathcal{X}'_\text{ph}\Delta_\text{ph} \right]^2}{a^2\Delta'^2_\text{ph}}.
\label{eq:eta}
\end{gather}
Here we note that the subscript ``$\text{ph}$" indicates that the above equations should be evaluated at the photon orbit, i.e. $r=r_\text{ph}$. With this information in hand, we can use the critical impact parameters $\xi$ and $\eta$ given by Eqs. (\ref{eq:xi}) and (\ref{eq:eta}) to study the contour of our BH shadow.

To investigate the effect of DM spike on the shadow images of our BH we shall  assume that the observer is located at some position given by the coordinates $(r_o,\theta_o)$, with $r_o$ being the distance of the observer and $\theta_o$ is the angular coordinate on observer's sky. Let us now introduce the following  coordinates on the observer's screen, $X$ and $Y$, defined by the relations \citep{000}
\begin{equation}
X=-r_o \frac{p^{(\phi)}}{p^{(t)}},\,\,\,\,Y=r_o\frac{p^{(\theta)}}{p^{(t)}},
\end{equation}
with $(p^{(t)},p^{(r)},p^{(\theta)},p^{(\phi)})$ being the tetrad components of the photon momentum with respect to locally non-rotating reference frame. It follows that, in the observer bases $e^{\mu}_{(\nu)}$ can be expanded in the coordinate bases 
\begin{eqnarray}\notag
p^{(t)}&=&-e^{\mu}_{(t)}p_{\mu}=E\,\zeta-\gamma\, J,\,\, p^{(\phi)}=e^{\mu}_{(\phi)}p_{\mu}=\frac{J}{\sqrt{g_{\phi\phi }}},\\
p^{(\theta )}&=& e^{\mu}_{(\theta)}p_{\mu}=\frac{p_{\theta }}{\sqrt{g_{\theta \theta}}},\,\,p^{(r)}=e^{\mu}_{(r)}p_{\mu}=\frac{p_r}{\sqrt{g_{rr }}},
\end{eqnarray}
where $E =-p_t$ and $p_{\phi}=J$, as we already noted, are conserved due to the associated Killing vectors. Furthermore, if we use  $p_{\theta}=\pm \sqrt{\Theta(\theta)}$, we can rewrite the celestial coordinates in terms of $\xi$ and $\eta$, as follows 
\begin{align}\nonumber
& X = -r_o \dfrac{\xi}{\sqrt{g_{\phi \phi}}(\zeta-\beta \xi)}|_{(r_o,\theta_o)},\\\label{beta}
& Y = \pm r_o \frac{\sqrt{\eta+a^2 \cos^2\theta-\xi^2 \cot^2\theta}}{\sqrt{g_{\theta \theta}}(\zeta-\beta \xi)}|_{(r_o,\theta_o)},
\end{align}
where
\begin{equation}
    \zeta=\sqrt{\frac{g_{\phi\phi}}{g_{t \phi}^2-g_{tt}g_{\phi \phi}}},
\end{equation}
and
\begin{equation}
    \beta=-\frac{g_{t \phi}}{g_{\phi \phi}}\zeta.
\end{equation}
 Finally to simplify the problem further we are going to consider that our observer is located in the equatorial plane ($\theta=\pi/2$) and very large but finite $r_o\sim 8.3$ kpc, we find
\begin{align}
& X = -\sqrt{f(r_o)}\, \xi,\\
& Y = \pm \sqrt{f(r_o)} \,\sqrt{\eta}.
\end{align}
 
  \begin{figure*}
 \centering
   \includegraphics[scale=0.55]{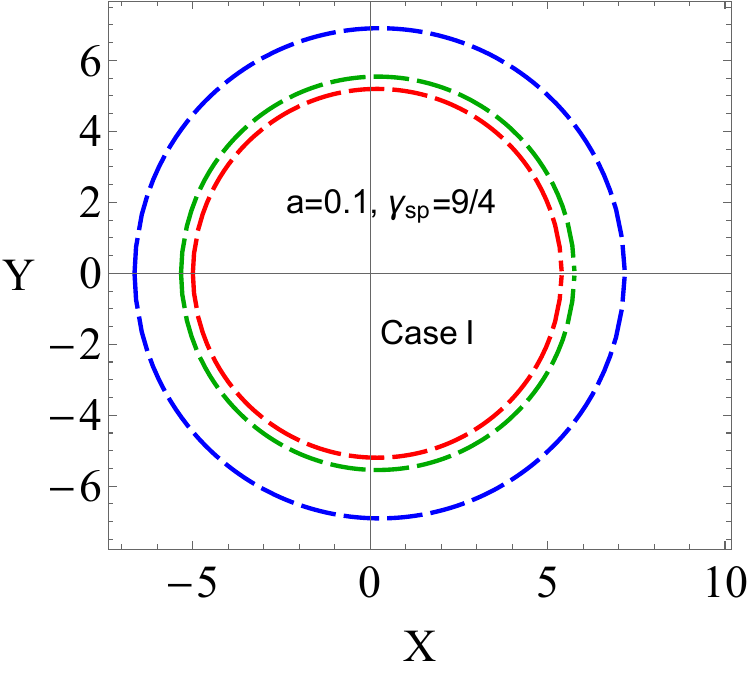}
     \includegraphics[scale=0.55]{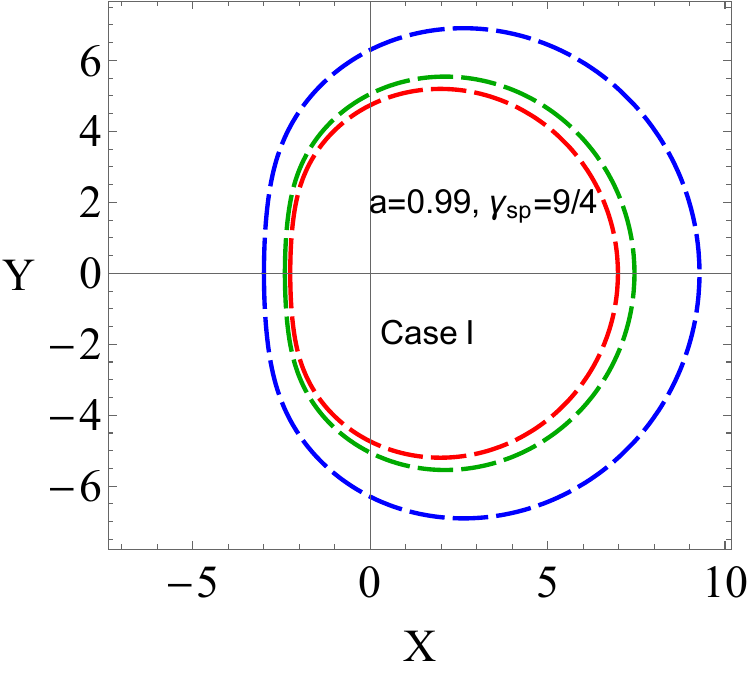}
    \includegraphics[scale=0.55]{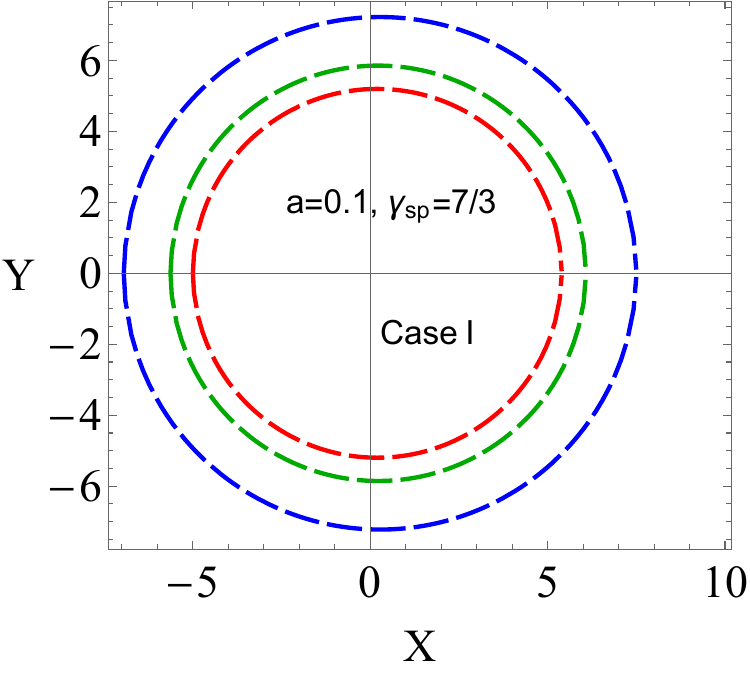}
   \includegraphics[scale=0.55]{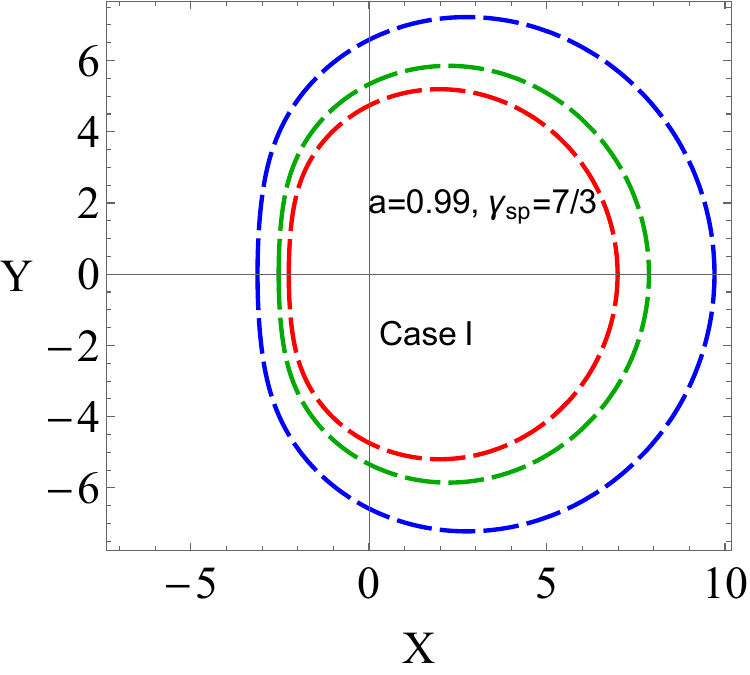}
    \caption{Upper panel: Shadows of the Sgr A$^\star$ BH surrounded by DM using Case I $(\omega=0)$ for $\gamma_\text{sp}=9/4$,  $R_\text{sp}\sim 0.91 $ kpc, and $\rho_\text{sp} \sim 1.39 \times 10^{-24}$g/cm$^3$ (red dashed curves), $\rho=10^4 \rho_\text{sp}$ (green dashed curves) and $\rho=10^5 \rho_\text{sp}$ (blue dashed curves). Here we have set the BH mass $M_\text{BH}=1$. Lower panel: Shadows of the Sgr A$^\star$ BH surrounded by DM using Case I for $\gamma_\text{sp}=7/3$,  $\rho_\text{sp} \sim 8 \times 10^{-23}$g/cm$^3$ and $R_\text{sp}\sim 0.235 $ kpc (red dashed curves)  $\rho=10^4 \rho_\text{sp}$ (green dashed curves) and $\rho=10^5 \rho_\text{sp}$ (blue dashed curves). $r_\text{b}=4$ and the observer is located at $r_o=10^{10}$, both measured in units of BH mass.}\label{shadow1}
\end{figure*}

\begin{deluxetable*}{lcccc}
\tablecaption{Typical shadow radii and angular sizes for Case I and Case II, respectively. We have set $a_*=0.65$ (see, \citep{Dokuchaev:2013xda}), $\gamma_\text{sp}=9/4$, $r_o=8.3$ kpc and $r_\text{b}=4$ in the units of the BH mass.}
\tablehead{\colhead{{$\rho_{sp}$  [g/cm$^3$ ]}} & \colhead{{$R^{I}_{sh}[M_{BH}$}]} & \colhead{{$\theta^{I}_s$[$\mu as$]}} & \colhead{{$R^{II}_{sh}[M_{BH}$]}} & \colhead{$\theta^{II}_s$[$\mu as$]}}
\startdata
$0$  &  5.052288582  & 51.67360723   & 5.052288582  & 51.67360723  \\
$1.39 \times 10^{-27}$  & 5.052288617 & 51.67360762   & 5.052288658  & 51.67360800  \\
$1.39 \times 10^{-24}$  & 5.052326161 & 51.67399160  & 5.052363780 & 51.67437633  \\
$1.39 \times 10^{-21}$  &  5.089451883  & 52.05370465  & 5.125837465 & 52.42584772 \\
$1.39 \times 10^{-20}$ & 5.389971481 & 55.12734771  &  5.664841181 & 57.93864969 \\
$1.39 \times 10^{-19}$  & 6.720201799  & 68.73262734  & 7.034594744  & 71.94816369  \\
\\
\enddata
\end{deluxetable*}

\label{tab:shadows1}
\begin{figure*}
 \centering
 \includegraphics[scale=0.55]{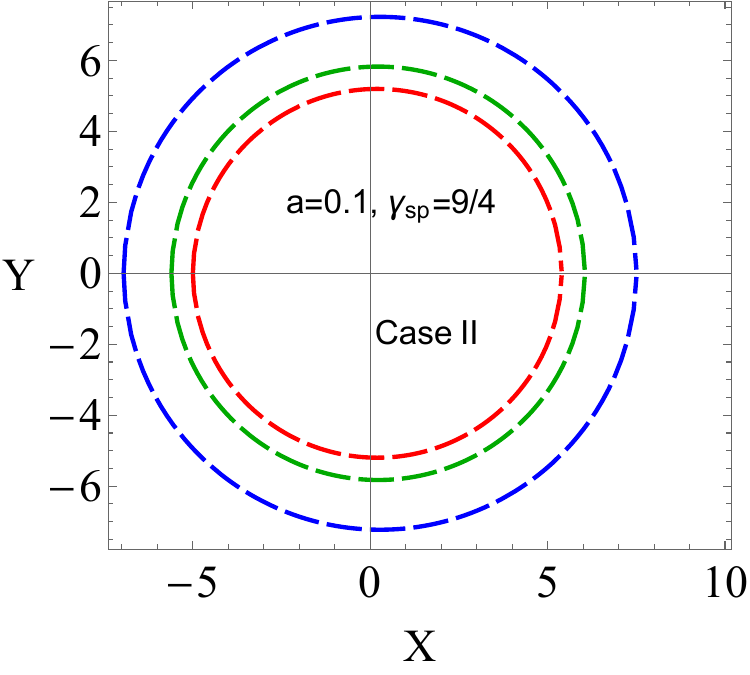}
     \includegraphics[scale=0.55]{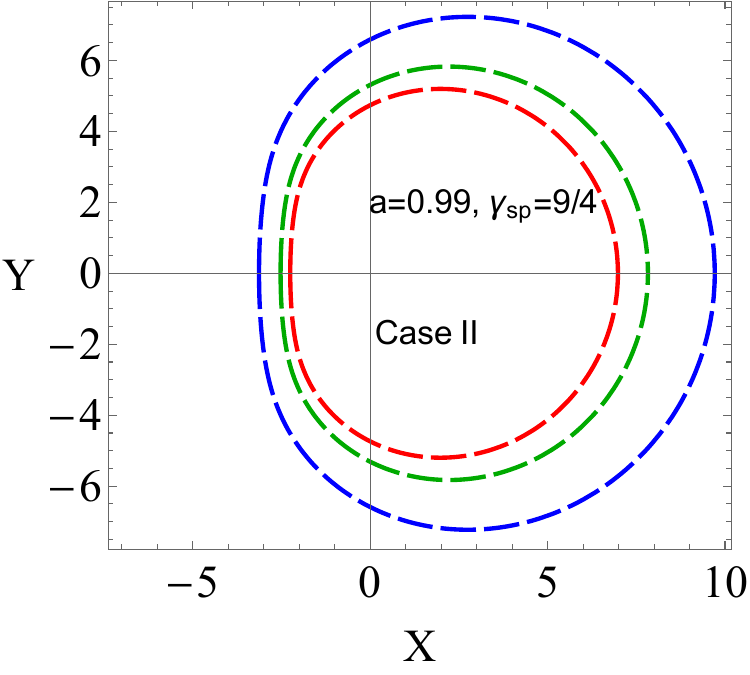}
    \includegraphics[scale=0.55]{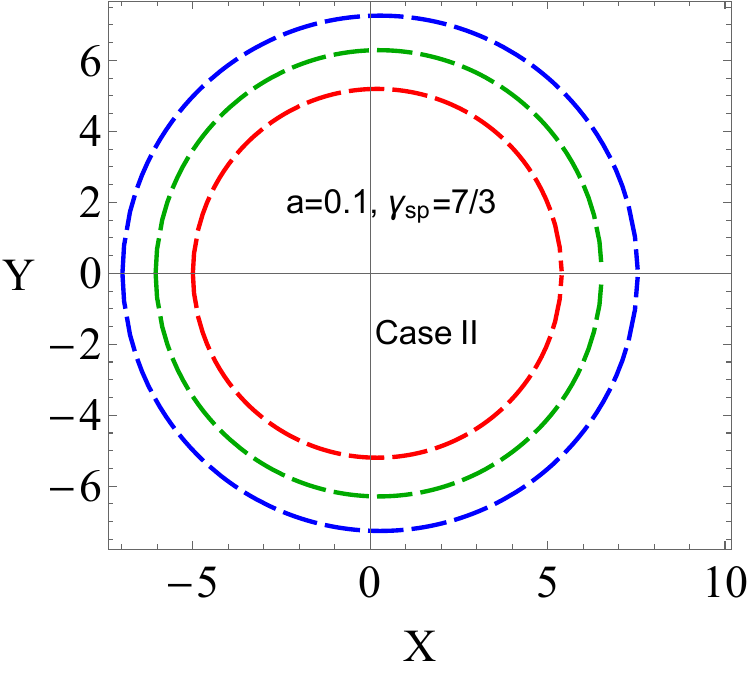}
   \includegraphics[scale=0.55]{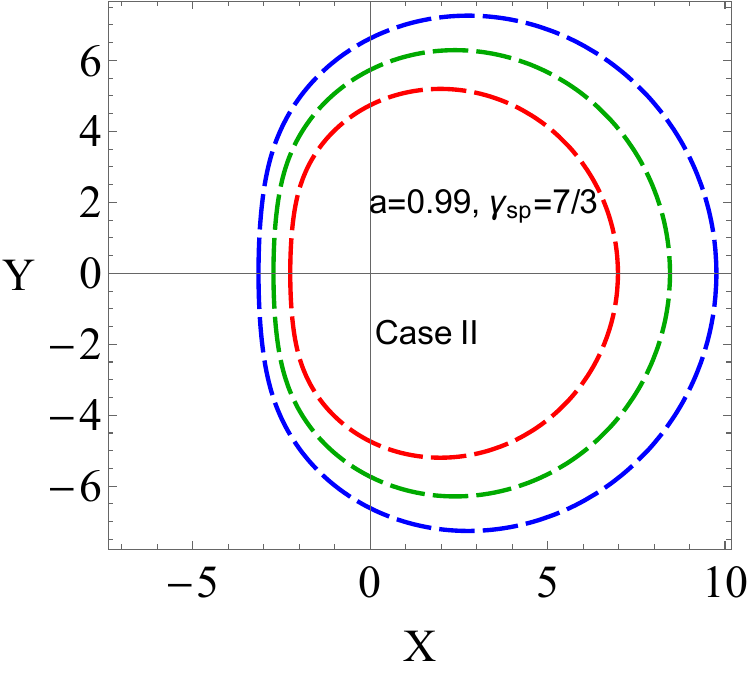}
    \caption{Upper panel: Shadows of the Sgr A$^\star$ BH surrounded by DM using Case II $(\omega=1)$ for $\gamma_\text{sp}=9/4$, $R_\text{sp}\sim 0.91 $ kpc, and $\rho_\text{sp} \sim 1.39 \times 10^{-24}$g/cm$^3$  (red dashed curves), $\rho=10^4 \rho_\text{sp}$ (green dashed curves) and $\rho=10^5 \rho_\text{sp}$ (blue dashed curves). Lower panel: Shadows of the Sgr A$^\star$ BH surrounded by DM using the realistic Case II for $\gamma_\text{sp}=7/3$,  $\rho_\text{sp} \sim 8 \times 10^{-23}$g/cm$^3$ and $R_\text{sp}\sim 0.235 $ kpc (red dashed curves), $\rho=10^4 \rho_\text{sp}$ (green dashed curves) and $\rho=10^5 \rho_\text{sp}$ (blue dashed curves). $r_\text{b}=4$ and the observer is located at $r_o=10^{10}$, both measured in units of BH mass.}\label{shadow2}
\end{figure*}

\begin{deluxetable*}{lcccc}
\tablecaption{Same as Tab.~\ref{tab:shadows1} but for $\gamma_\text{sp}=7/3$.}
\tablehead{\colhead{{$\rho_{sp}$  [g/cm$^3$ ]}} & \colhead{{$R^{I}_{sh}[M_{BH}$}]} & \colhead{{$\theta^{I}_s$[$\mu as$]}} & \colhead{{$R^{II}_{sh}[M_{BH}$]}} & \colhead{$\theta^{II}_s$[$\mu as$]}}
\startdata
$0$  &  5.052288582  & 51.67360723   & 5.052288582  & 51.67360723  \\
$8 \times 10^{-27}$  & 5.052288617 & 51.67360762  & 5.052288658 & 51.67360800  \\
$8 \times 10^{-23}$  & 5.052680498 & 51.67761564  & 5.053072414  & 51.68162406  \\
$8 \times 10^{-21}$  & 5.091033350  & 52.06987953  & 5.128903765 & 52.45720913 \\
$8 \times 10^{-20}$  & 5.403134906  & 55.26198010   & 5.686394456 & 58.15909147  \\
$8 \times 10^{-19}$  & 6.754301013 & 69.08138598 & 7.061280088  & 72.22109501 \\
\enddata

\end{deluxetable*}
\label{tab:shadows2}


 Working in the case with $\theta_0=\pi/2$, we shall use the definition adopted in Refs.~\citep{Zhang:2019glo,Feng:2019zzn} where the typical shadow radius is defined in terms of the leftmost and rightmost coordinates $r_A$ and $r_\text{B}$, and reads
\begin{equation}
R_\text{sh}=\frac{1}{2}\left(X^{+}-X^{-}\right),
\end{equation}
along with the condition $Y(r=r_A)=Y(r=r_\text{B})=0$.  Following Ref. \citep{Jusufi:2020dhz}, we do a simple algebra from the last equation which results in an expression for the typical shadow radius
\begin{equation}
R_\text{sh}=\frac{\sqrt{2\,f(r_o)}}{2}\left(\sqrt{\frac{ r_\text{ph}^{+}}{f'(r)|^\text{Kerr}_{r_\text{ph}^{+}}}}+\sqrt{\frac{ r_\text{ph}^{-}}{f'(r)|^\text{Kerr}_{r_\text{ph}^{-}}}}\right).
\end{equation}
Note that inside $r<r_\text{b}$, the spacetime metric is a pure Kerr BH and the photon orbit is not affected at all by the DM. However, the shadow radius is modified due to the nontrivial topology of the surrounding DM. In other words, the motion of photons from the BH to an observer located at $r_o$ will be affected by the surrounding geometry. For instance, defining $r_\text{\text{ph}}^{\pm}$ as the radius of circular null geodesics for the spinning BH, in the special case when $a=0$, we obtain $r_{\text{ph}}^{\pm}=3M_{BH}$, yielding shadow radius radius for the static metric as $R_{sh}=\sqrt{f(r_o)} 3 \sqrt{3} M_{BH}$. 

In order to study the realistic effect of DM on the shadow radius we need to use the DM mass function $M(r)$ and study the motion of light in this region near the BH. Here we shall consider two scenarios as before, viz., $\gamma_\text{sp}=9/4$ and $\gamma_\text{sp}=7/3$. To investigate a realistic model supported by the observational data for the the Sgr A$^\star$  BH, we notice that for $\gamma_\text{sp}=9/4$, we have  $\rho_\text{sp} \sim 1.39 \times 10^{-24}$g/cm$^3$ and $R_{sp}\simeq 0.91 $ kpc, while for $\gamma_\text{sp}=7/3$, we have  $\rho_\text{sp} \simeq 8 \times 10^{-23}$g/cm$^3$ and $R_\text{sp}\simeq 0.235 $ kpc. Apart from these astrophysically motivated values, we will also analyze the effect of increasing the DM density on the observables. During this analysis though, the DM densities cannot be arbitrarily large, as we will see now. 

When exploring the parameter space, it is very important to ensure that the spacetime remains free of pathologies. A classic example of this is the Kerr metric itself, where the spin parameter $a_*$ is bounded between $1$ and $-1$ to ensure the singularity always remains covered by the event horizon. Some of the conditions that must be satisfied everywhere outside the horizon include
\begin{itemize}
    \item The metric determinant must always be negative.
    \item $g_{\phi\phi}$ is always greater then $0$.
    \item $g_{rr}$ remains finite.
\end{itemize}
Imposing these conditions on the metric in the static case, we get the following condition
\begin{equation}
    g(r) > 0
\end{equation}
everywhere outside the horizon, while the condition in the rotating case is
\begin{equation}
    \Delta(r)/H(r) > 0.
\end{equation}
Though there are no analytical solutions for these equations, they can easily be evaluated numerically with scientific computing softwares like Mathematica. For instance, for a specific $R_\text{sp}$ and $\gamma_\text{sp}$ (note that we have fixed $r_\text{b} = 4$), we can get bounds on $\rho_\text{sp}$. In particular, for the two cases considered in this work, we get
\begin{eqnarray}\label{eq:rholimits}
    \gamma_\text{sp} = 9/4,\; R_\text{sp} &=& 0.91 \text{ kpc}:\nonumber\\ \rho_\text{sp} &<& 5.34\times 10^{-19} \text{g}/\text{cm}^3\\
     \gamma_\text{sp} = 7/3,\; R_\text{sp} &=& 0.235 \text{ kpc}:\nonumber\\ \rho_\text{sp} &<& 2.37\times 10^{-18} \text{g}/\text{cm}^3
\end{eqnarray}
in the static case. The bounds remain nearly identical in the rotating case. Changing the value of $r_\text{b}$, though, does affect the bounds on $\rho_\text{sp}$. 

The BH shadows for a variety of cases are plotted in Figs.~\ref{shadow1} and~\ref{shadow2}. Several interesting features can be observed in these figures. Firstly, the shadows remains nearly identical to their no-DM counterparts for realistic DM parameters (plotted with red dashed lines). With the increase of the DM density $\rho_\text{sp}$, the shadow radius slowly increases. The effect on the shadow size becomes significant only when the DM density increases by order of $10^4-10^5$. We also note that the effect of DM on the shadow radius/angular diameter is slightly stronger in the case $\gamma_\text{sp}=7/3$ compared to $\gamma_\text{sp}=9/4$. The observable quantity such as the angular diameter of the Sgr A$^{*}$ BH can be estimated using the observable $R_{sh}$ as follows 
\begin{equation}
	\theta_s = 2R_\text{sh} M_{\text{BH}}/D,
\end{equation} 
where $M_{BH}$ is the BH mass and $D$ is the distance between the BH and the observer. Or, alternatively, we can express this relation as 
\begin{equation}
	\theta_s = 2 \times 9.87098 \times 10^{-6} R_s\frac{M}{M\textsubscript{\(\odot\)}}\frac{1\text{kpc}}{D}\, \mu \text{as}.
\end{equation}
Tabs.~\ref{tab:shadows1} and~\ref{tab:shadows2} show these quantities for a range of DM densities, and we can see that a significant change of shadow size, say, larger then the $17\%$ uncertainty reported by the EHT collaboration in 2017~\citep{eht}, occurs only for the largest DM density. 


\section{\label{sec:images}Radiative signature of a BH in a dark matter spike}
\begin{figure*}
\centering
     \includegraphics[scale=0.45]{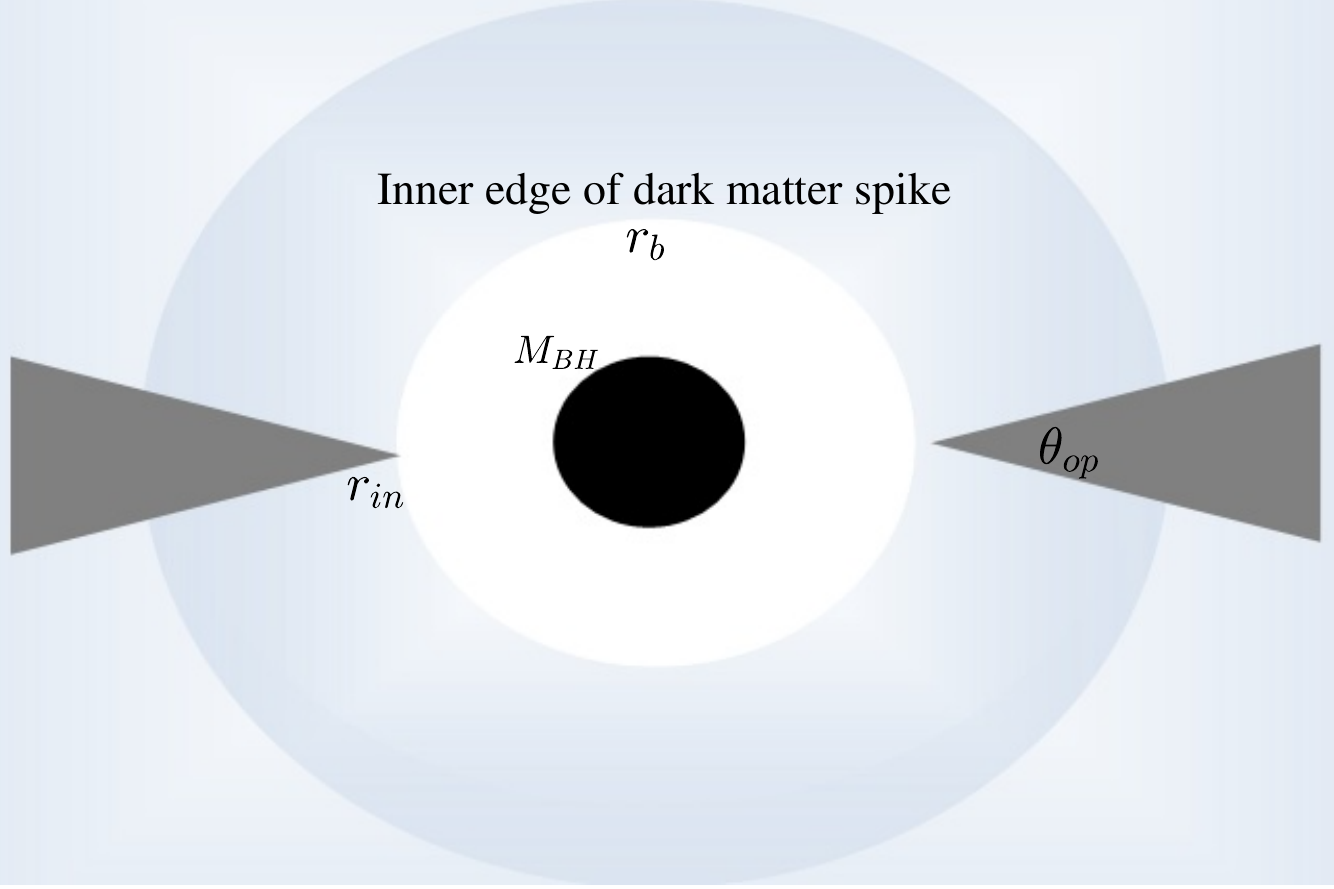}
    \caption{\label{fig:disk} Sketch of the geometrically-thick accretion disk model immersed in a dark matter spike. The central black circle indicates the BH, and the disk is shown in gray. $r_{\textrm{in}}$ and $\theta_{\textrm{op}}$ mark the location of the inner edge and opening angle of the disk, respectively. The observer plane is shown highly amplified and placed almost equatorial in inclination. Red lines indicate photon trajectories and red shaded regions illustrate that part of the trajectories which lies inside the disk.}
\end{figure*}

\begin{figure*}
\centering
    \includegraphics[width=\columnwidth]{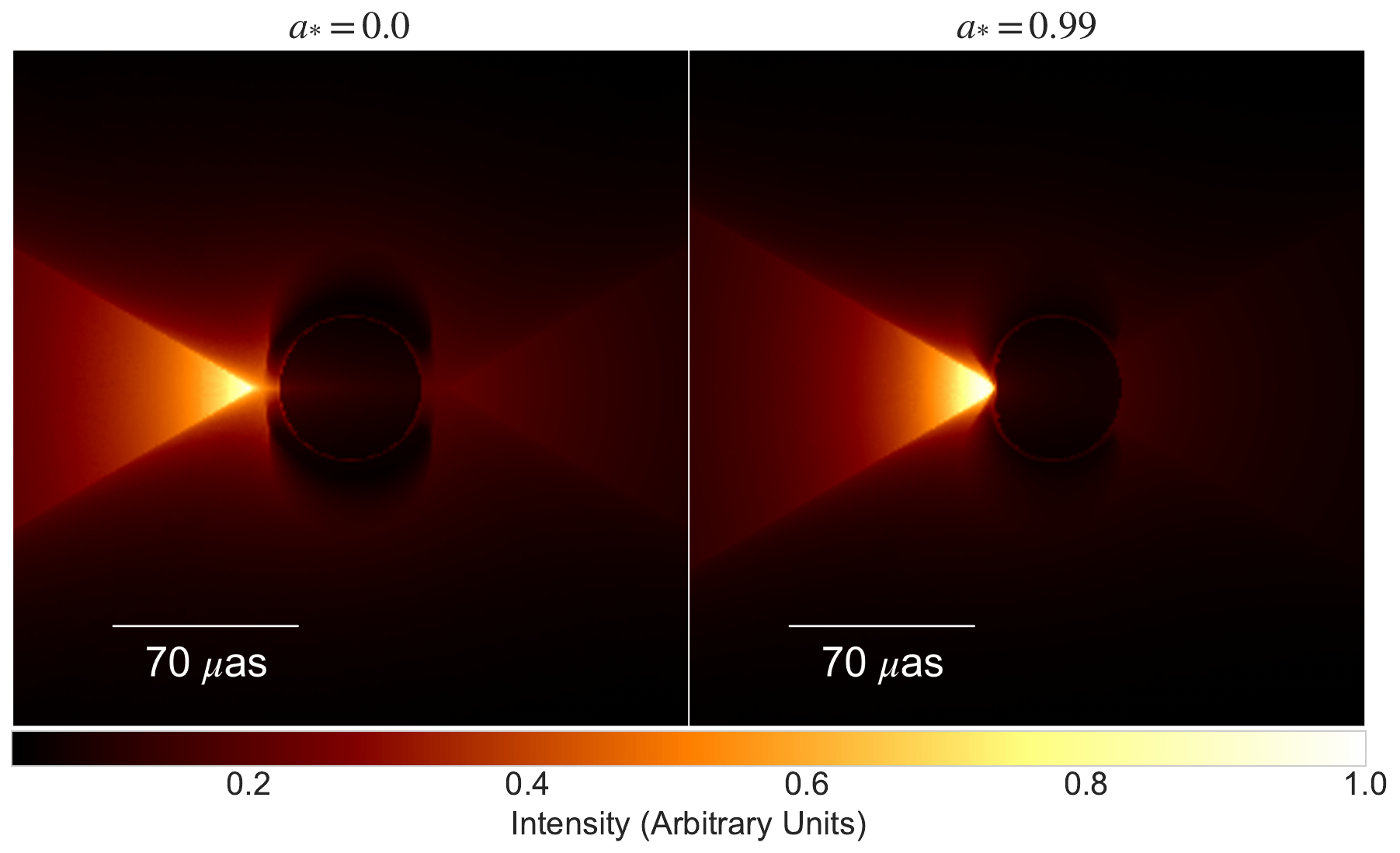}
    \includegraphics[width=\columnwidth]{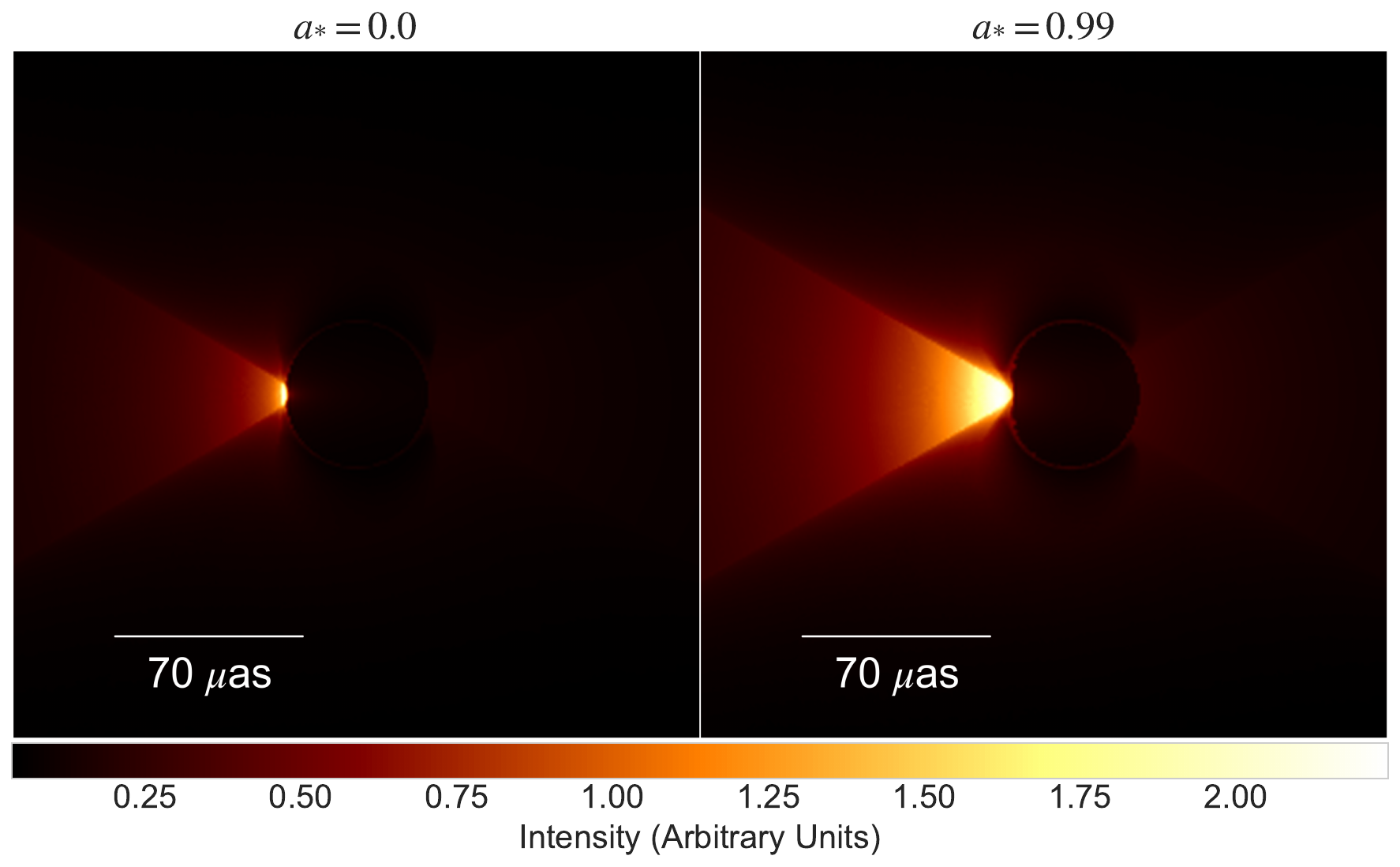}
    \caption{\label{fig:im-nodm} Images of a BH surrounded by a geometrically-thick optically-thin disk, radiating as a monochromatic power-law, and without any DM. In the panels on the left the disk inner edge is located at the ISCO ($r_{\textrm{in}} = r_{\textrm{ISCO}}$) and in the panels on the right, it is at the equatorial photon orbit ($r_{\textrm{in}} = r_{\textrm{ph}}$). See the text for more details.}
\end{figure*}
Continuing our exploration of the effect of the DM spike on the observables from the central region of the Milky Way, in this section, we compute the effect of the DM spike on the \textit{image} of the BH. We consider a toy model for the accretion disk and the radiation profile, with the aim to identify qualitative effects of the DM spike. To this end, we make several assumptions about the emitting region and the emission mechanism around the hole. The model used here is based on the one used in Ref.~\citep{Nampalliwar:2020asd}. We consider a very simple accretion model of a geometrically thick disk, inspired from Ref.~\citep{Vincent:2020dij} and illustrated in Fig.~\ref{fig:disk}. The disk extends from some inner radius $r_{\textrm{in}}$ (which we fix at either the innermost stable circular orbit $r_{\textrm{ISCO}}$ or the equatorial photon orbit $r_{\textrm{ph}}$) to some outer radius (effectively imposed by the radial fall-off of the intensity, see below), and has an opening angle of $\theta_{\textrm{op}}$ which we set to $30^{\circ}$. The disk is assumed to be axisymmetric relative to the $z$-axis, which coincides with the BH spin axis. Matter in the disk flows in circular orbits independent of the height from the equator. Thus, a particle located at some $(r,\theta)$ on the disk has the same velocity as a particle on the equatorial plane at a distance $r\sin{\theta}$.
In all cases, we assume the emitting region is optically thin.

The calculation of the intensity of the emitting region requires some assumption about the radiative processes and emission mechanisms. Generically, the observed specific intensity $I_{\mathrm{obs}}$ at the observed photon frequency $\nu_{\mathrm{obs}}$ at the point $(X,Y)$ of the observer's image plane is given by 
\begin{equation}
    I_{\mathrm{obs}`}(\nu_{\mathrm{obs}},X,Y) = \int_{\gamma}\mathrm{g}^3 j(\nu_{e})dl_{\mathrm{prop}},  
\end{equation}
where $\mathrm{g} = \nu_{\mathrm{obs}}/\nu_{e}$ is the redshift factor, $\nu_{e}$ is the photon frequency as measured in the rest-frame of the emitter, $j(\nu_{e})$ is the emissivity per unit volume in the rest-frame of the emitter, and $dl_{\mathrm{prop}} = k_{\alpha}u^{\alpha}_{e}$ is the infinitesimal proper length as measured in the rest-frame of the emitter. The redshift factor is evaluated from 
\begin{equation}
    \mathrm{g} = \frac{k_{\alpha}u^{\alpha}_{\mathrm{obs}}}{k_{\beta}u^{\beta}_{e}} 
\end{equation}
where $k^{\mu}$ is the four-velocity of the photons, $u^{\alpha}_{e}$ is the four-velocity of the accreting material, $u^{\mu}_{\mathrm{obs}}$ = $(1,0,0,0)$ is the four-velocity of the observer and $\lambda$ is an affine parameter along the photon path. 
For specific emissivity, we assume a simple model in which the emission is monochromatic, with emitter's rest-frame frequency $\nu_{\star}$, and has the following power law profile:
\begin{equation}
    j(\nu_{e}) \propto \frac{\delta(\nu_{e}-\nu_{\star})}{r^2},
\end{equation}
where $\delta$ is the Dirac delta function. 
Integrating the intensity over all the observed frequencies, we obtain the observed flux
\begin{equation}\label{eq:flux}
    F_{obs}(X,Y) \propto -\int_{\gamma} \frac{\mathrm{g}^3 k_t}{r^2k^r}dr.
\end{equation}

To create the image, we begin with placing an observer at some large distance $r_o$ such that
\begin{equation}
    r_\text{b} \ll r_o \sim   R_{\textrm{sp}}.
\end{equation}
We define a coordinate system $(X, Y, Z)$ centered at that point on the observer's screen where a radial vector from the center of the BH intersects the screen, with the $Z$-axis lying along this radial vector. On the screen, we place 300 cells each in the $X$ and the $Y$ direction (from 0 to $40M$). Photons are traced backwards in time from the observer's screen until one of the following conditions is satisfied: either the photon crosses the event horizon, or it escapes to infinity (i.e., the photon's radial coordinate becomes larger then $r_o$). For the portion of the photon trajectory inside the disk (illustrated in Fig.~\ref{fig:disk} with red shaded regions), radiative flux is accumulated using Eq.~\ref{eq:flux}. The discrete flux map thus generated is normalized, by dividing the flux at each point by the total flux, and interpolated to produce a smooth image.

\begin{figure}
\centering
    \includegraphics[width=\columnwidth]{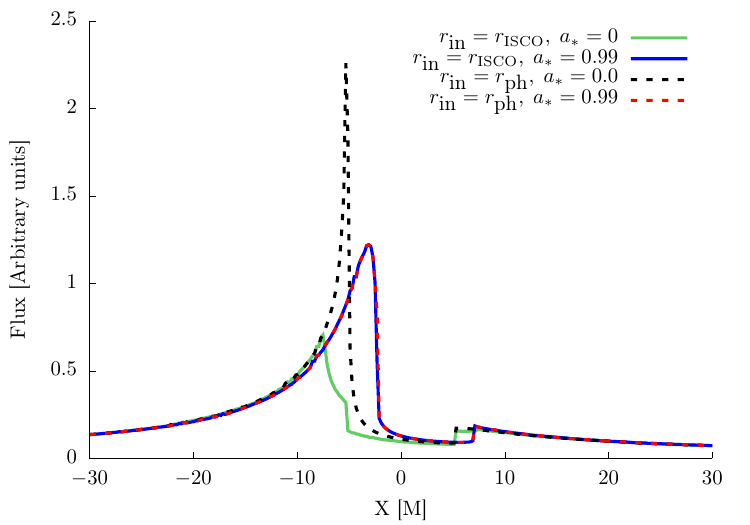}
    \caption{\label{fig:flux-nodm} Radiative flux along the $X$-axis for the same four configurations shown in Fig.~\ref{fig:im-nodm}. See the text for more details.}
\end{figure}

\begin{figure*}
\centering
    \includegraphics[scale=0.42]{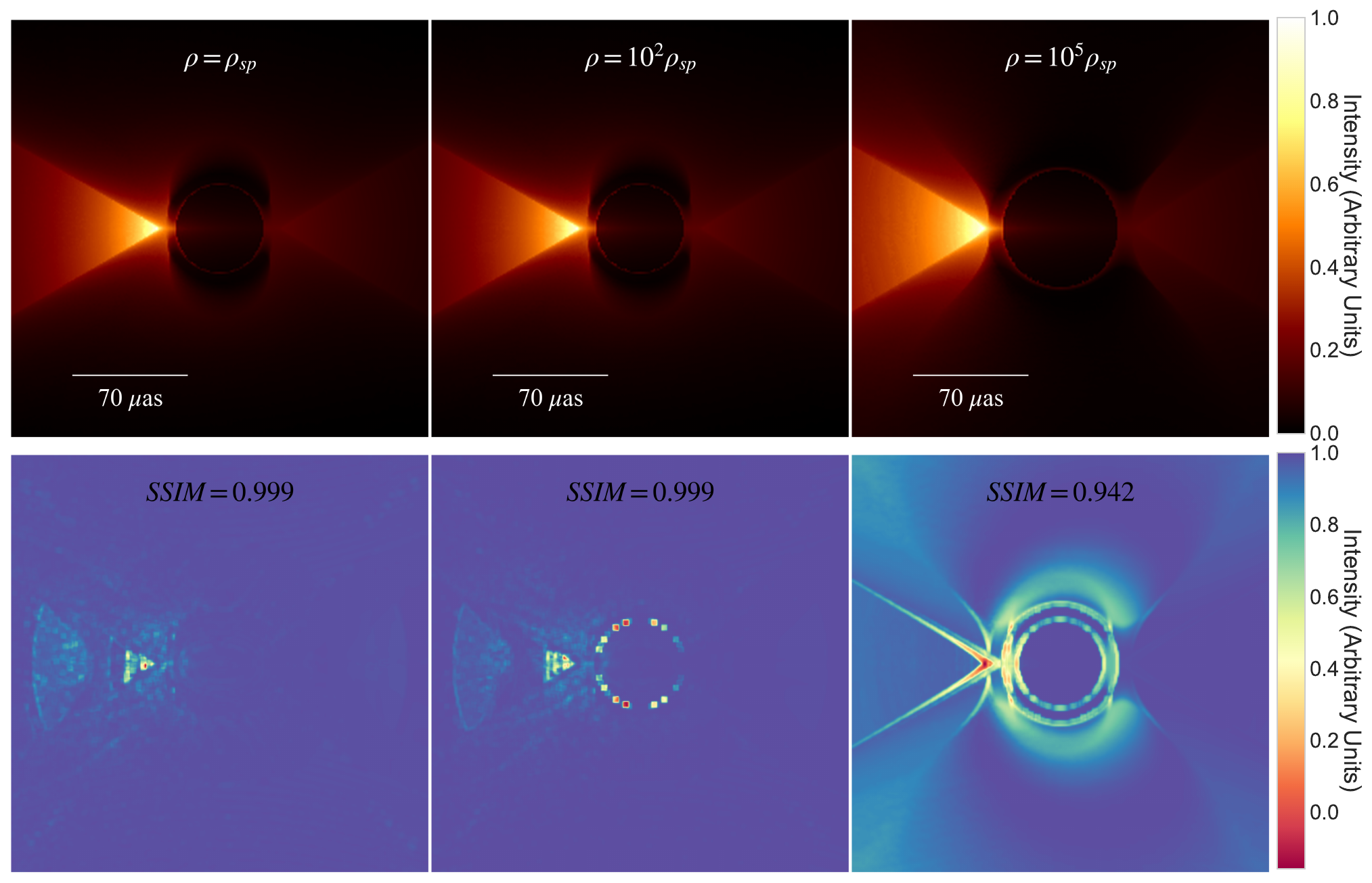}
    \caption{\label{fig:im-case1-isco}Images of a BH for $a_*=0.0$, $\gamma_\text{sp}=9/4$, $R_\text{sp}=0.91$~kpc, and the inner edge of the disk lying at the innermost stable circular orbit ($r_{\textrm{in}} = r_{\textrm{ISCO}}$). The upper row shows the images for different DM densities labeled in the figure, and the lower row shows the difference of the images in upper row with those for same $a_*$ and $r_{\textrm{in}}$ but without the effect of DM, as seen in the first image in the left panel in Fig.~\ref{fig:im-nodm}. See the text for more details.}
\end{figure*}

\begin{figure*}
\centering
    \includegraphics[scale=0.42]{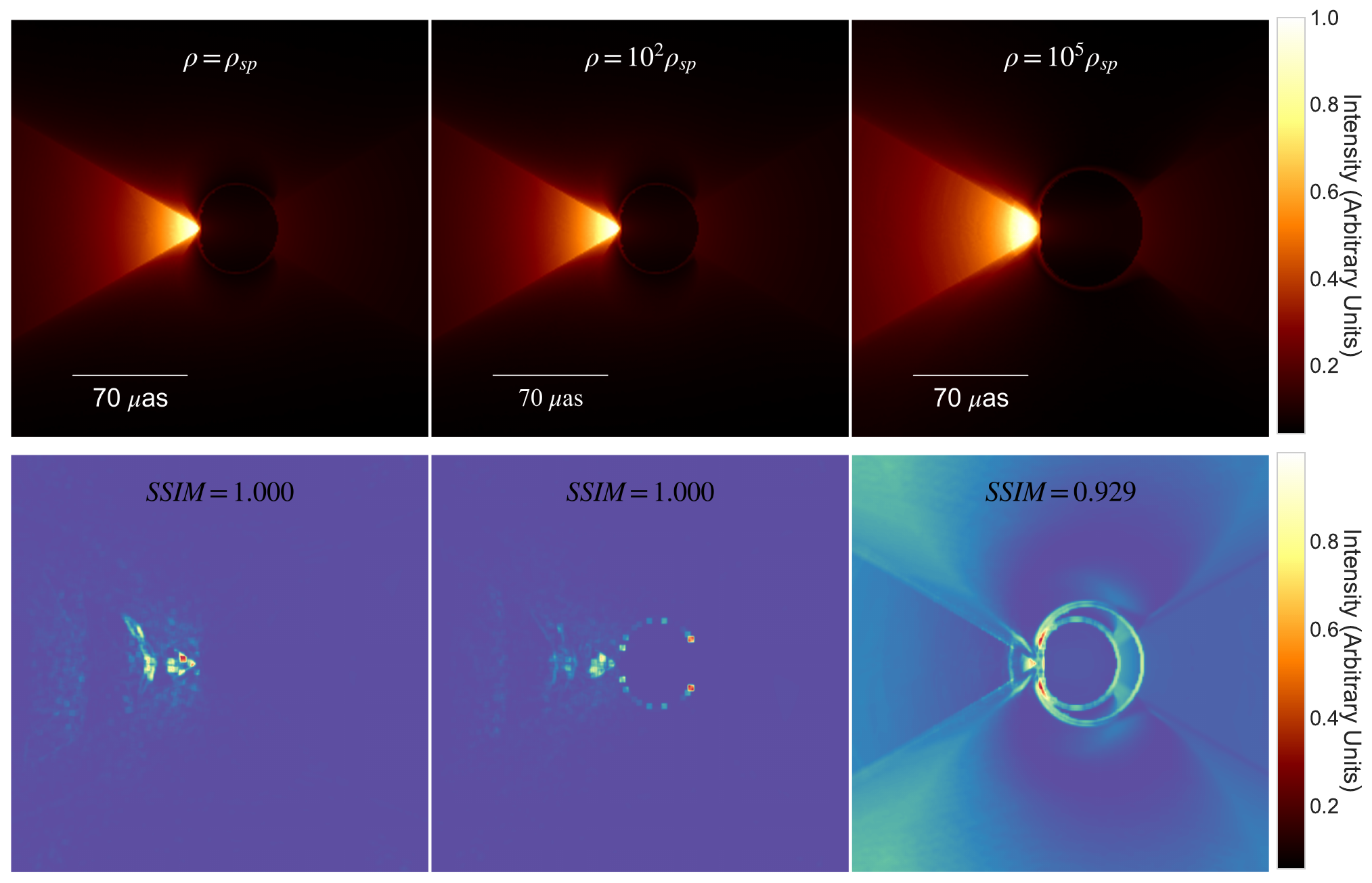}
    \caption{\label{fig:im-case2-isco}Images of a BH for $a_*=0.99$, $\gamma_\text{sp}=9/4$, $R_\text{sp}=0.91$~kpc, and the inner edge of the disk lying at the innermost stable circular orbit ($r_{\textrm{in}} = r_{\textrm{ISCO}}$). The upper row shows the images for different DM densities labeled in the figure, and the lower row shows the difference of the images in upper row with those for same $a_*$ and $r_{\textrm{in}}$ but without the effect of DM, as seen in the second image in the left panel in Fig.~\ref{fig:im-nodm}.}
\end{figure*}

\begin{figure*}
\centering
    \includegraphics[scale=0.42]{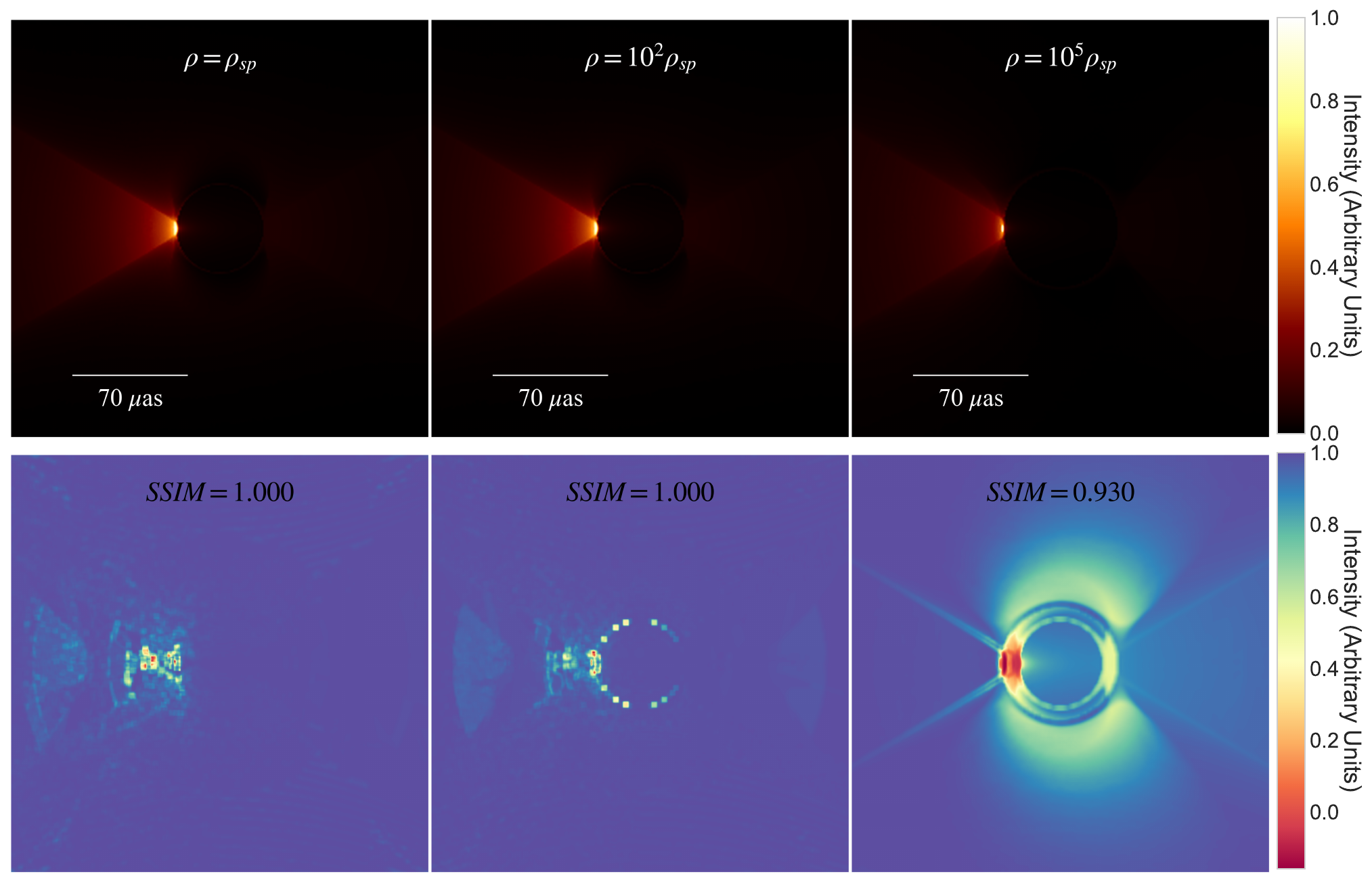}
    \caption{\label{fig:im-case1-ph}Images of a BH for $a_*=0.0$, $\gamma_\text{sp}=9/4$, $R_\text{sp}=0.91$~kpc, and the inner edge of the disk lying at the photon orbit ($r_{\textrm{in}} = r_{\textrm{ph}}$). The upper row shows the images for different DM densities labeled in the figure, and the lower row shows the difference of the images in upper row with those for same $a_*$ and $r_{\textrm{in}}$ but without the effect of DM, as seen in the first image in the right panel in Fig.~\ref{fig:im-nodm}. See the text for more details.}
\end{figure*}

\begin{figure*}
\centering
    \includegraphics[scale=0.42]{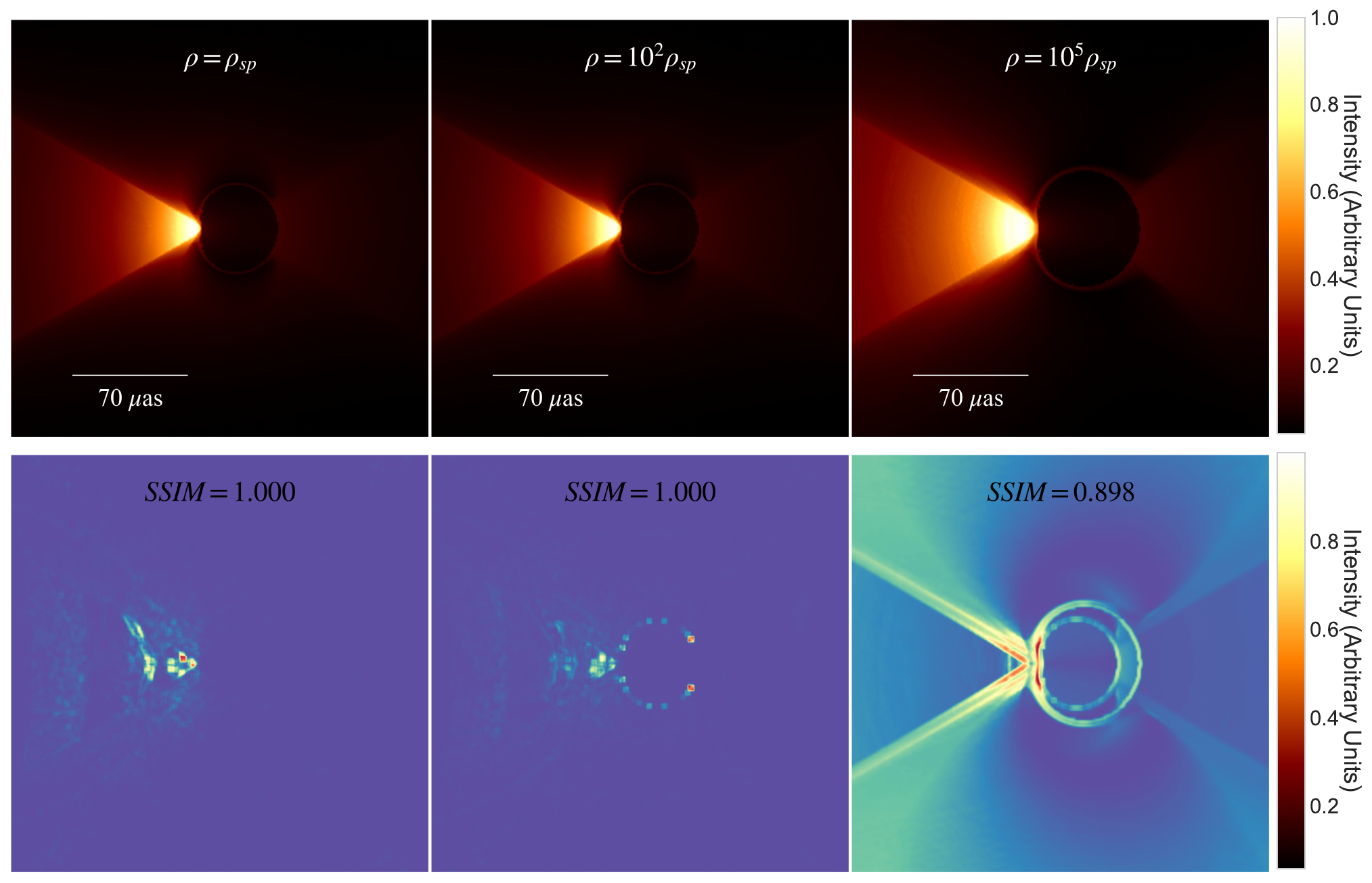}
    \caption{\label{fig:im-case2-ph}Images of a BH for $a_*=0.99$, $\gamma_\text{sp}=9/4$, $R_\text{sp}=0.91$~kpc, and the inner edge of the disk lying at the photon orbit ($r_{\textrm{in}} = r_{\textrm{ph}}$). The upper row shows the images for different DM densities labeled in the figure, and the lower row shows the difference of the images in upper row with those for same $a_*$ and $r_{\textrm{in}}$ but without the effect of DM, as seen in the second image in the right panel in Fig.~\ref{fig:im-nodm}.}
\end{figure*}

The images for several different scenarios are presented in Figs.~\ref{fig:im-nodm} to~\ref{fig:im-case2-ph}, and we discuss them in succession. We begin with Fig.~\ref{fig:im-nodm}, which is made assuming no DM spike (i.e., a Kerr background). This serves as a benchmark when we discuss images with DM below. In the panels on the left in Fig.~\ref{fig:im-nodm}, we plot the images for two different spins with the disk inner edge fixed at the respective ISCOs (i.e., $r_{\textrm{in}} = r_{\textrm{ISCO}}$). Several interesting features can be inferred from these. Firstly, there is a distinct ring which coincides with the shadows that were discussed in the previous section. This is a very important feature because among all possible measurable quantities from a BH image, the shadow is least affected by the (often not-so-well understood) accretion disk microphysics and is a direct indicator of the BH geometry. Secondly, the cross-sectional shape of the specific disk model we chose is reflected in the image as a bright funnel, brightest at the narrow end  (cf. Fig.~\ref{fig:disk}). Thirdly, photons are strongly redshifted (blueshifted) closer to hole, resulting in a dimming (brightening) in the intensity. Since in the static case the ISCO (at $6M$) and the photon orbit (at $3M$) are far apart, there is a gap between the bright narrow end and the shadow ring. In the higher spin case, since the ISCO (at $1.45M$) is closer to the photon orbit (at $1.17M$), the bright narrow end is also closer to the shadow ring. In fact, this leads to a very similar image when the inner edge is fixed at the photon orbit ($r_{\textrm{in}} = r_{\textrm{ph}}$), for the higher spin case, as shown in the panels on the right in Fig.~\ref{fig:im-nodm}. For the static case, understandably, the image is quite distinct when switching from $r_{\textrm{in}} = r_{\textrm{ISCO}}$ to $r_{\textrm{in}} = r_{\textrm{ph}}$. One, rather obvious, feature is that the narrow end of the bright funnel is closer to the shadow ring. A second, more interesting, feature is the much sharper dimming in the funnel-like region close to the narrow end, even when compared to the high spin cases. This is related to the facts that, in the static case, photons coming from regions close to the photon orbit suffer much higher blueshift then those coming from regions close to the ISCO, and frame-dragging reduces the blueshift in the spinning case compared to the static case. This is shown explicitly in Fig.~\ref{fig:flux-nodm}, where we plot the flux along the X-axis for the above four images.  While intriguing, it is crucial to note here that the $r_{\textrm{in}} = r_{\textrm{ph}}$ case, following our model assumptions, has matter flowing in circular orbits even inside the ISCO, which is an inherently unstable scenario.  


 Armed with an understanding of the various features of the images constructed with our model, we proceed to analyzing the effect of the DM spike. In order to keep the analysis succinct, we only present the results for the DM profile given by Case I of Sec.~\ref{sec:profile} and for $\gamma_\text{sp} = 9/4$. (Results for Case II and results with $\gamma_\text{sp} = 7/3$ exhibit qualitatively similar features.) In particular, the DM acts as a pressure-less fluid, the metric is given by Eq.~\ref{eq:le} in the static case and by Eq.~\ref{eq:rotmet} in the rotating case, and the metric functions $g(r)$, $f(r)$ and $H(r)$ are given in Eqs.~\ref{eq:gfunc},~\ref{eq:ffunc} and~\ref{eq:hfunc}, respectively. Figs.~\ref{fig:im-case1-isco} and~\ref{fig:im-case2-isco} show the images for $r_{\textrm{in}} = r_{\textrm{ISCO}}$ for two different spins, respectively, and a range of DM densities. The base DM density, referring to the leftmost panel in the top row in each figure, is the realistic value discussed in Sec.~\ref{sec:shadows} and, for $\gamma_\text{sp} = 9/4$, equals $\rho_{sp} = 1.39\times10^{-24}$ g/cm$^3$. 
 Comparing these panels with the no DM scenario of Fig.~\ref{fig:im-nodm}, we see that the DM spike has a negligible effect on the images for realistic DM densities. The subsequent panels in the top rows show the images for larger DM densities. Even when enhanced a hundred times, the with-DM images look quite similar to their no-DM counterparts. Only when the densities are enhanced to very high values do the images acquire an unambiguously distinct look. Similar trends are seen for images with $r_{\textrm{in}} = r_{\textrm{ph}}$, presented in Figs.~\ref{fig:im-case1-ph} and~\ref{fig:im-case2-ph}. 
 
 While a visual inspection does give a qualitative idea, we can try to quantify the difference between the images, especially those that appear distinct. To this end, we use a quality metric which is a human visual-perception metric known as structural similarity index (SSIM) \citep{Zhou:2006ssim}. Defining a pair of images as \textit{I} and \textit{K}, the SSIM is defined as 
 \begin{equation}
     \text{SSIM}({\textit{I}, \textit{K}}) = \mathcal{L}(\textit{I}, \textit{K}) \mathcal{S}(\textit{I}, \textit{K})\mathcal{C}(\textit{I}, \textit{K})
     \label{eq:ssim}
 \end{equation}
where $\mathcal{L}(\textit{I}, \textit{K})$ is the luminance with the structure $\mathcal{S}(\textit{I}, \textit{K})$ and the contrast $\mathcal{C}(\textit{I}, \textit{K})$. These terms are further dependent on the average flux of each of the image $\mu_{{\textit{I}, \textit{K}}}$, the variance $\sigma^2_{\textit{I}, \textit{K}}$ of the individual image and the covariance of the image pair  $\sigma_{{\textit{I}\textit{K}}}$. Hence Eq. \ref{eq:ssim} can be rewritten as 
\begin{equation}
    \text{SSIM}({\textit{I}, \textit{K}}) = \left( \frac{2\mu_{\textit{I}}\mu_{\textit{K}}}{\mu_{\textit{I}}^2 + \mu_{\textit{K}}^2} \right) \left( \frac{2\sigma_{\textit{I}}\sigma_{\textit{K}}}{\sigma_{\textit{I}}^2 + \sigma_{\textit{K}}^2} \right)\left( \frac{2\sigma_{\textit{I}\textit{K}}}{\sigma_{\textit{I}}\sigma_{\textit{K}}} \right)
\end{equation}
where
\begin{eqnarray}
\mu_{\textit{I}} &=& \Sigma^{N}_{i=1}{\frac{\textit{I}_i}{N}} \\
\sigma^2_{\textit{I}} &=& \frac{\Sigma^{N}_{j=1}(\textit{I}_j - \mu_j)^2}{(N-1)} \\
\sigma_{\textit{I}\textit{K}} &=& \frac{\Sigma^{N}_{j=1}(\textit{I}_j - \mu_\textit{I})(\textit{K}_j - \mu_\textit{K})}{(N-1)}
\end{eqnarray}
Two identical images would have SSIM $=1$, and dissimilarities reduce the SSIM value. In the lower rows of Figs.~\ref{fig:im-case1-isco}--\ref{fig:im-case2-ph}, we present the pixel-by-pixel difference between the with-DM and no-DM images (Fig.~\ref{fig:im-case1-isco} and the leftmost image in Fig.~\ref{fig:im-nodm}, Fig.~\ref{fig:im-case2-isco} and the second from left image in Fig.~\ref{fig:im-nodm}, and so on) and the corresponding value of SSIM. The pixel-by-pixel difference confirms our visual inference: only very high DM densities have noticeably different images. The SSIM quantifies this inference, and is nearly equal to $1$ for all cases except when the DM densities are very high.

Is this high DM density effect, even if unrealistic, detectable? We explore this question in two ways. First, in the rest of this section, we analyze its degeneracy with other \textit{source}-related quantities that affect the image. Then, in the following section, we analyze the effect of the \textit{instrument} on the detectability.

We have seen that the image is affected by various parameters of our model, which can be classified as: astrophysical ($r_{\textrm{in}}$), GR ($a_*$) and DM ($\rho_{sp}$). For simplicity, we use only one parameter from each category. In an astrophysical scenario, their true values are often unknown and, if their effects on the image are degenerate, this can weaken the inference about detectability drawn up to now. The SSIM-based analysis is a powerful tool~\citep{Mizuno:2018lxz} and can be used to explore this potential degeneracy to some extent. Since the astrophysical parameter is strongly model-dependent (a realistic accretion scenario will almost certainly not have a sharp inner edge as considered here) and takes unrealistic values ($r_{\textrm{in}} = r_{\textrm{ph}}$ leads to an inherently unstable disk in the inner regions), we limit our exploration to degeneracies between the GR and the DM parameters, keeping the astrophysical parameter fixed. Let us first consider the $r_{\textrm{in}} = r_{\textrm{ISCO}}$ set of images. Starting with the ($a_* = 0, \rho = \rho_{sp}$) image, we either change $a_*$ to $0.99$ (and calculate the SSIM) or $\rho$ to $10^5\rho_{sp}$ (and calculate the SSIM). Comparing these two SSIMs with the SSIM calculated between ($a_* = 0.99, \rho = \rho_{sp}$) and ($a_* = 0, \rho = 10^5\rho_{sp}$) gives us an indication whether the effects of these parameters on the image are degenerate. I.e., we define
\begin{eqnarray}
    A &:& a_* = 0,\; \rho = \rho_{sp},\\
    B &:& a_* = 0.99,\; \rho = \rho_{sp},\\
    C &:& a_* = 0,\; \rho = 10^5\rho_{sp},
\end{eqnarray}
then, if 
\begin{eqnarray}
    \mathrm{SSIM}(C,B) &<& \mathrm{SSIM}(B,A),\\
    \mathrm{SSIM}(C,B) &<& \mathrm{SSIM}(C,A),
\end{eqnarray}
we can say that the effects of the GR and the DM parameters can be distinguished in the image. A similar comparison is possible with 
\begin{eqnarray}
    A &:& a_* = 0.99,\; \rho = 10^5\rho_{sp},
\end{eqnarray}  
while keeping $B$ and $C$ as before. Tab.~\ref{tab:ssim-inf} presents these SSIM values in the first two rows. For both choices of image $A$, the above condition is satisfied, showing that very high DM densities imprint a strong, \textit{and unique}, signature on the BH image. A similar analysis can be done with the $r_{\textrm{in}} = r_{\textrm{ph}}$ set of images, and the SSIM values are presented in the lower two rows of Tab.~\ref{tab:ssim-inf}. Here, we see that the above condition is again satisfied for one choice of image $A$. For the other choice, viz., $a_* = 0.99, \rho = 10^5\rho_{sp}$, both SSIM$(C,A)$ and SSIM$(C,B)$ are very small, which suggests that the spin $=0$ images (corresponding to image $C$ and shown in Fig.~\ref{fig:im-case1-ph}) are quite distinct from the spin $=0.99$ images (shown in Fig.~\ref{fig:im-case2-ph}). 

It is important to keep in mind the limitations of the analysis performed here. The biggest limitation is the astrophysical modelling: the accretion and radiation features are certainly more complex around Sgr A$^\star$ then considered here~\citep{Yuan:2014gma}. The mass of the central BH is not known exactly, which will introduce additional uncertainties and, possibly, degeneracies. We kept the DM parameters like $R_\text{sp}$ and $r_\text{b}$ constant, whereas in an actual analysis, these will have to be simultaneously determined with the DM density. 

\begin{deluxetable}{lccc}
\tablecaption{A comparison of SSIM values for different pairs of infinite-resolution images. The images are denoted by their $(r_{\textrm{in}}, a_*, \rho)$ values. For the top two rows, image $B$: $(r_{\textrm{ISCO}}, 0.99, 0)$ and image $C$: $(r_{\textrm{ISCO}}, 0, 10^5\rho_{sp})$, whereas for the bottom two rows, image $B$: $(r_{\textrm{ph}}, 0.99, 0)$ and image $C$: $(r_{\textrm{ph}}, 0, 10^5\rho_{sp})$.}
\tablehead{\colhead{{Image $A$}} & \colhead{{SSIM($B,A$)}} & \colhead{{SSIM($C,A$)}} & \colhead{{SSIM($C,B$)}}}
\startdata
$r_{\textrm{ISCO}},\; 0,\; 0$ & $0.905$ & $0.99$ & $0.839$ \\
$r_{\textrm{ISCO}},\; 0.99,\; 10^5\rho_{sp}$ & $0.939$ & $0.934$ & $0.839$ \\
$r_{\textrm{ph}},\; 0,\; 0$  & $0.786$ & $0.968$ & $0.534$  \\
$r_{\textrm{ph}},\; 0.99,\; 10^5\rho_{sp}$  & $0.946$ & $0.445$ & $0.534$\\
\enddata
\end{deluxetable}
\label{tab:ssim-inf}

\section{\label{sec:radio}Detectability of the dark matter spike with Radio Interferometry}

\begin{deluxetable}{cccc}
\tablecaption{Locations of sites in the Event Horizon Telescope like array configuration considered in this work.}
\tablehead{\colhead{\textbf{Site}} & \colhead{\textbf{Lat.~(\degr)}} & \colhead{\textbf{Lon.~(\degr)}}& \colhead{\textbf{SEFD. (Jy)}}}
\startdata
ALMA & -23.03 & -67.75 & 90    \\
APEX & -23.01 & -67.76 & 3500  \\
BAJA  & 30.87  & -115.46 & 10000 \\
BOL   & -16.25 & -68.13  & 10000 \\
CARMA & 37.1   & -118.14 & 10000 \\
DRAK  & -29.3  & 29.27   & 10000 \\
GAM   & 23.25  & 16.17   & 10000 \\
HAY   & 42.43  & -71.49  & 2500  \\
JCMT & 19.82 & -155.48 & 6000  \\
KAUAI & 21.79  & -159.51 & 10000 \\
KEN   & -0.15  & 37.31   & 10000 \\
KP & 31.96 & -111.61 & 10000 \\
LMT & 18.98 & -97.31 & 600   \\
PIKES & 38.65  & -105.04 & 10000 \\
PDB   & 44.44  & 5.91    & 1500  \\
PV    & 36.88  & -3.39   & 1400  \\
SMA & 19.82 & -155.48 & 4900  \\
SMT & 32.70 & -109.89 & 5000  \\
SPT & -90.00 & 45.00 & 5000  \\
VLT   & -24.48 & -70.4   & 10000 \\
\enddata
\end{deluxetable}
\label{table:next}

\begin{figure}[t!]
\centering
\includegraphics[width=\columnwidth]{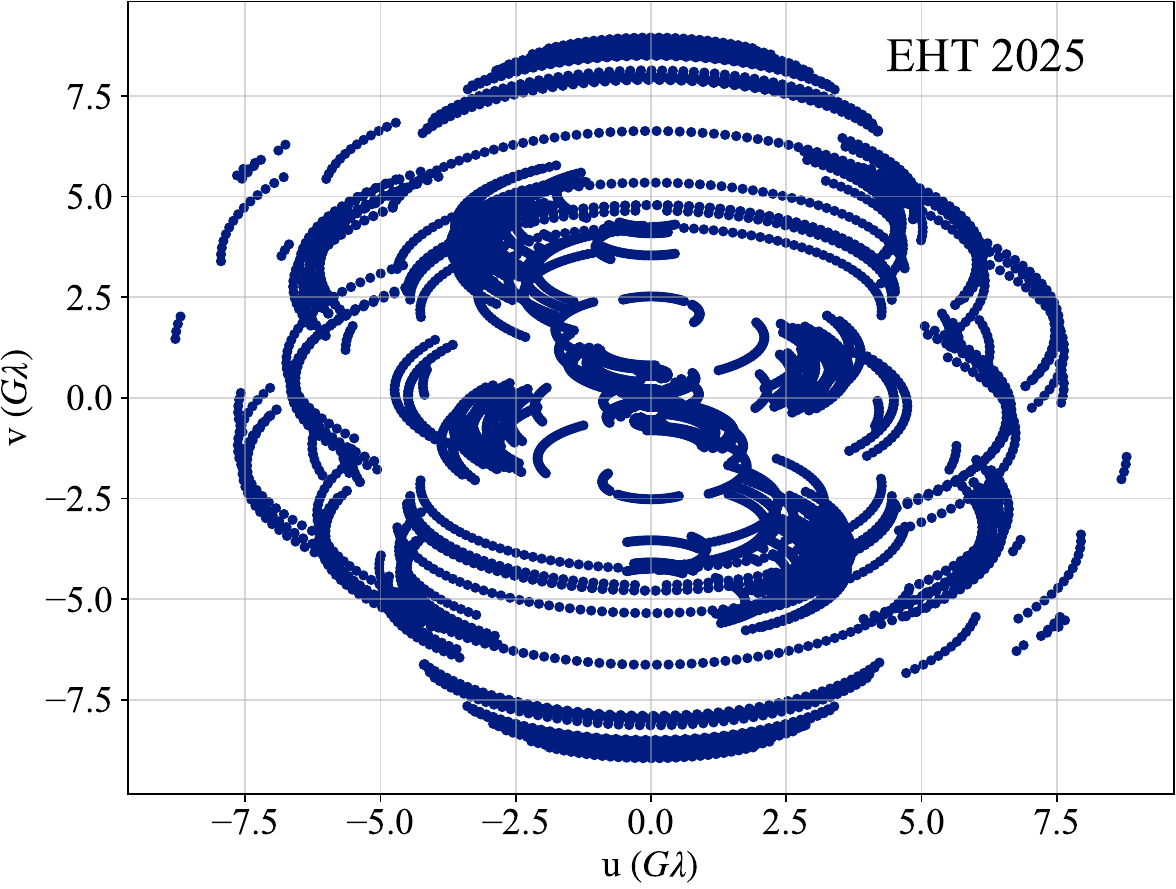}
\caption{Baseline coverage of Sgr A$^\star$ for the EHT-like array configuration mentioned in Tab.~\ref{table:next}.}
\label{fig:uv}
\end{figure}
\begin{figure*}
\centering
\includegraphics[scale=0.30]{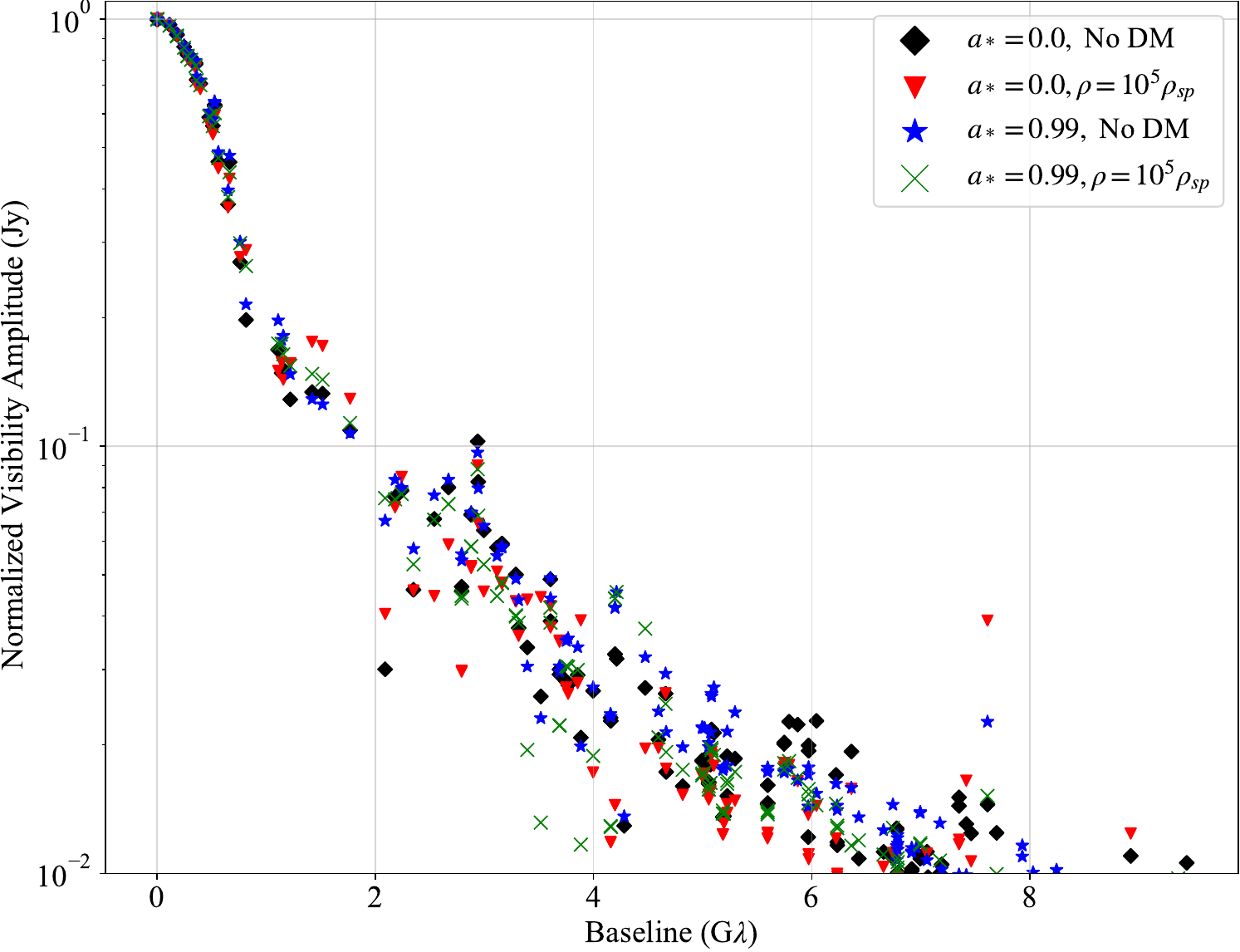}
\includegraphics[scale=0.30]{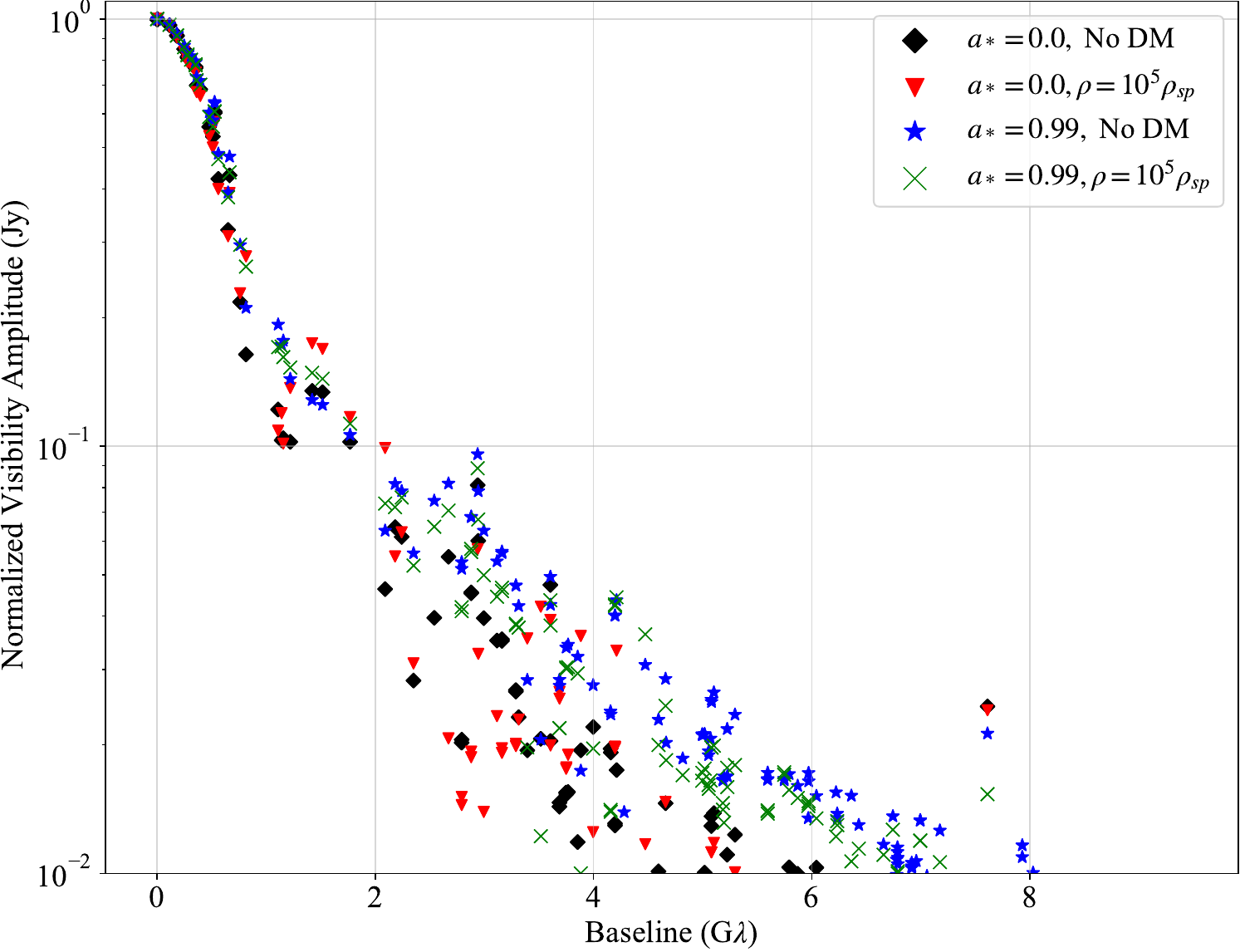}
\caption{Time-averaged normalized visibility amplitude profiles (log scale) for Sgr A$^\star$ for the simulation described in Sec.~\ref{sec:radio}, for two values of $r_{\textrm{in}}$: $r_{\textrm{ph}}$ (left panel) and $r_{\textrm{ISCO}}$ (right panel), and two values each of $a_*$ and $\rho$ (labeled in the plots). See the text for more details.}
\label{fig:vis_amp}
\end{figure*}

\begin{figure*}
\includegraphics[scale=0.41]{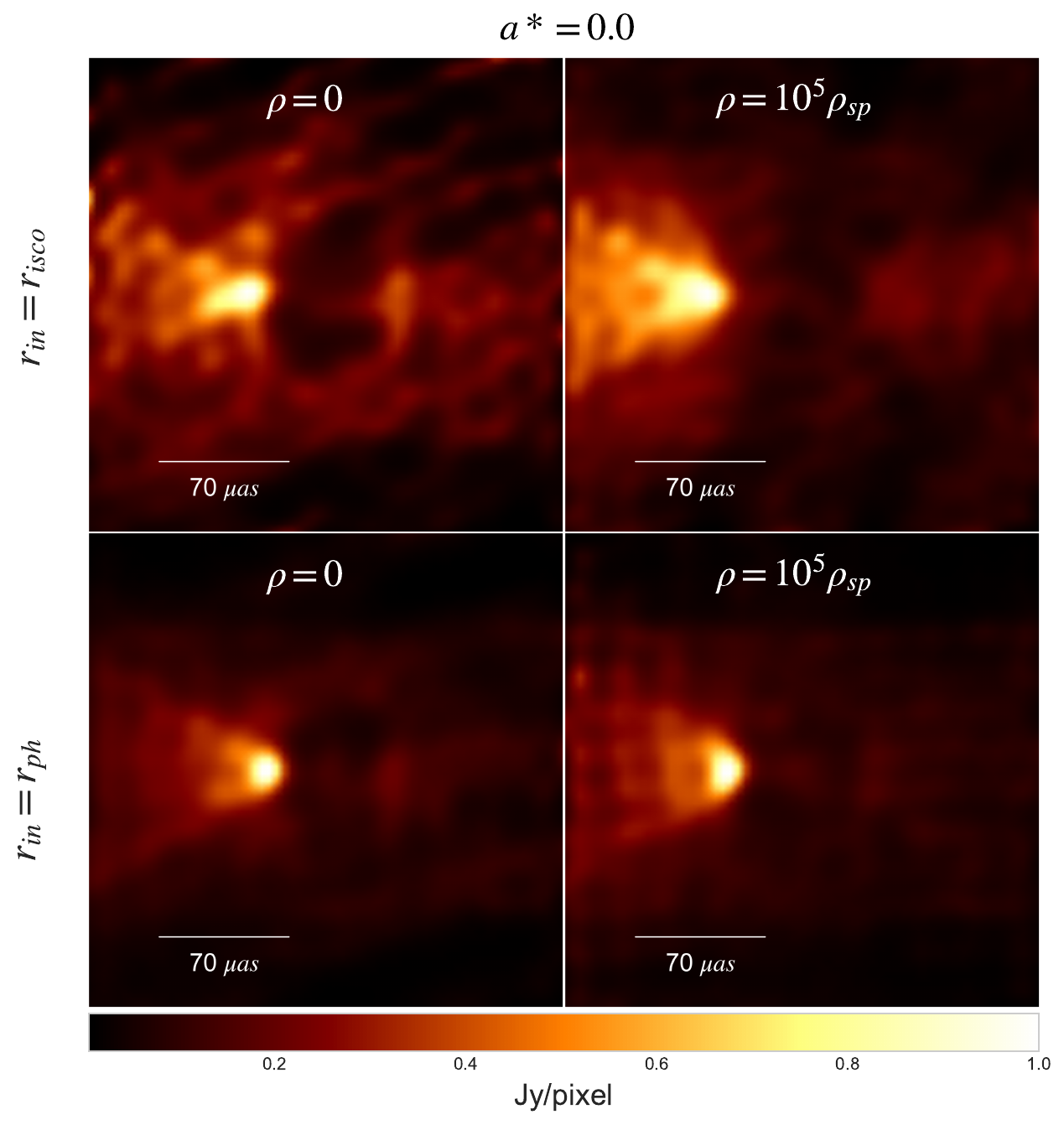}
\includegraphics[scale=0.41]{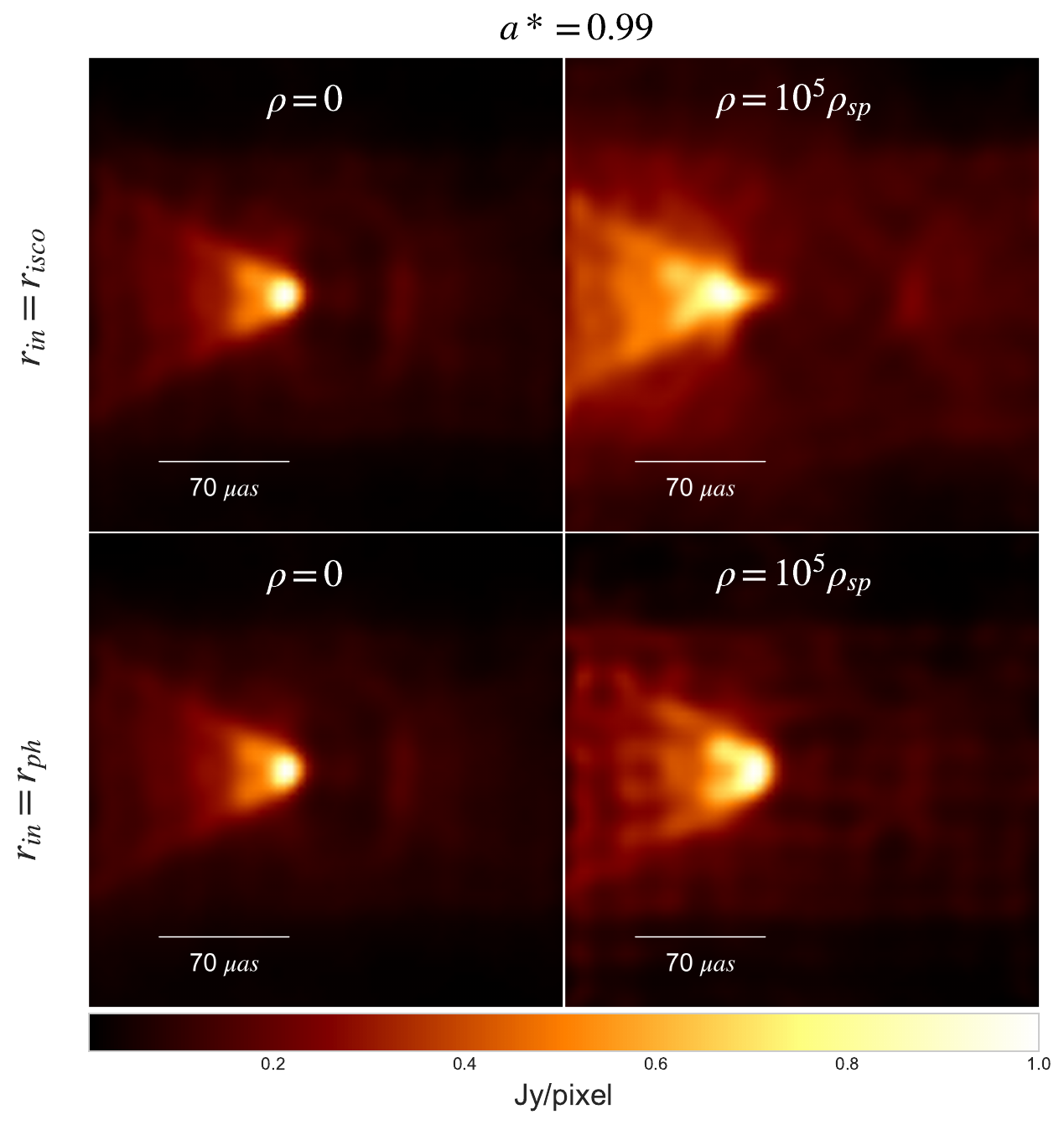}
\caption{Reconstructed images of Sgr A$^\star$ with the EHT-like array configuration mentioned in Tab.~\ref{table:next} and discussed in Sec.~\ref{sec:radio}. Left panel is for $a_*=0.0$ and right panel is for $a_*=0.99$.}
\label{fig:rec-images}
 \end{figure*}

In the above sections, we have analyzed imaging scenarios progressively closer to observations, starting with pure shadows in Sec.~\ref{sec:shadows} and looking at ray-traced snapshots in Sec.~\ref{sec:images}. The images analyzed in the previous section are \textit{infinite}-resolution, in the sense that a pristine source which experiences no interstellar matter on its way to a perfect detector, that can detect every photon that comes its way, will produce such images. In reality, there will be interstellar scattering, and the baselines will cover only part of the observing region. To mimic a more realistic situation, and explore the detectability of the DM spike, we simulate observations of Sgr A$^\star$ surrounded by a DM spike with an EHT-like antenna and analyze the simulations. 

Till the April 2018 observing campaign, the EHT had 8 telescopes for observing Sgr A$^\star$ \citep{Palumbo1:2019}: the Atacama Large (sub) Millimeter Array (ALMA), in Chile; the Atacama Pathfinder Experiment Telescope (APEX), also in Chile; the James Clark Maxwell Telescope (JCMT), near the summit of Mauna Kea in Hawaii; the Large Millimeter Telescope (LMT), in Mexico; the 30m telescope on Pico Veleta in Spain (PV); the Submillimeter Array (SMA), located near JCMT; the Submillimeter Telescope (SMT); located on Mount Graham in Arizona; and the South Pole Telescope (SPT), operating at the National Science Foundation's South Pole research station. The EHT also includes the Greenland Telescope but it cannot observe Sgr A$^\star$. 
For this particular analysis, we include some extra stations which are likely to join the future EHT observations~\citep{ngeht:2021}. All the sites used in our simulations are listed in Tab.~\ref{table:next}. In what follows, we refer to this array configuration as EHT-like, and perform our analysis with this configuration.

We start by defining the image brightness distribution function in the sky, $I(x,y;\mathbf{\Theta_p})$. The observable from radio interferometery is not the actual flux, rather, what is typically reported is the complex visibility, which is related to the Fourier transform of the flux via the van Cittert-Zernike theorem \citep{TMS}
\begin{equation}
    \mathcal{V}(u,v;\mathbf{\Theta_p}) = \int\int e^{-2\pi(xu+yv)}I(x,y;\mathbf{\Theta_p})dxdy
\label{eq:visibility}
\end{equation}
where $\mathbf{\Theta_p}$ is a tuple of parameters for the images produced in the previous section, ($u, v$) are the fourier components of the sky position coordinates, ($x, y$) are related to the the baseline vectors $\Vec{b}$ connecting every pair of antennas in the array, projected orthogonally to the line of sight to a particular source over the wavelength of the observation. Complex visibilities are measured at the baselines that are constructed by all the possible pairs of stations in an array.

We generate the synthetic radio images with \texttt{ehtim}~\citep{Chael:2018}, using the array configuration reported in Tab.~\ref{table:next} for the image reconstruction. The following parameters were used in the simulations: $\Delta \nu=4$ GHz bandwidth, $t=24$ hours, corresponding to a full day, at a central frequency of 230 GHz. This unusually long observation time allows us to present the ideal-case scenario, as the total observing time is one of the important parameters in the imaging process. The longer the observation time, the larger the ($u, v$) coverage allowing us to reconstruct a better image. With this setup, we take the following steps:
\begin{itemize}
    \item Calculate the ($u, v$) coverage for the source visibility using the given configuration while taking the sensitivity of the telescopes into account.
    \item Calculate the interferometric visibilities corresponding to the ($u, v$) grid.
    \item Include phase errors and noise.
    \item Use Regularised Maximum likelihood (RML) method to reconstruct static images from synthetic VLBI data.
\end{itemize}
Following this procedure, we perform the synthetic observations of the Galactic center. The visibility amplitudes are calculated by Fourier transforming the images and sampling them with the projected baselines of the array. During the simulations, we also include the effects of thermal noise and phase errors to mimic realistic observations. For a full day of observation, the EHT-like array has a sufficiently well sampled ($u, v$) coverage, as can be seen from Fig.~\ref{fig:uv}. This produces highly detailed visibility amplitude plots, as shown in Fig.~\ref{fig:vis_amp} (cf. Fig.~1 in Ref.~\citep{Akiyama:2019eap}).

The primary outcome of this analysis, for our purposes, is the reconstructed image. These images can be used to verify the inferences drawn from the infinite-resolution images in Sec.~\ref{sec:images}. We show the reconstructed images for a few scenarios in Fig.~\ref{fig:rec-images}, specifically, for spins $=0.0$ and $0.99$ (in the left and right panels, respectively), for $r_\text{in}=r_\text{ISCO}$ and  $r_\text{ph}$ (in the top and bottom panels, respectively), and for $\rho = 0$ and $10^5\rho_{sp}$ (in the left and right images in each panel, respectively). As in Sec.~\ref{sec:images}, we present the analysis for Case I of Sec.~\ref{sec:profile} and for $\gamma_\text{sp} = 9/4$, noting that the inferences drawn remain valid for Case II and for $\gamma_\text{sp} = 7/3$. Generally, we observe that the reconstructed images have lost the distinctive features that the infinite-resolution images had, to a large extent. For instance, the $(r_{\textrm{ph}}, 0.99, 0)$ image (bottom left in the right panel) and the $(r_{\textrm{ISCO}}, 0.99, 0)$ image (top left in the right panel) are quite similar both at infinite-resolution (SSIM $\sim 1$) and after reconstruction (SSIM $\sim 1$), but the former image and the $(r_{\textrm{ph}}, 0., 0)$ image (bottom left in the left panel), while highly distinct at infinite-resolution (SSIM $=0.786$), are nearly identical after reconstruction (SSIM $\sim 1$). Note that we are following the notation used in the previous section, denoting an image by its $(r_{\textrm{in}}, a_*, \rho)$ values. We perform an SSIM-based analysis, along the lines of Sec.~\ref{sec:images}, for the reconstructed images as well, and the results are reported in Tab.~\ref{tab:ssim-rec}. These SSIM values, all nearly equal to one, imply that, even in the ideal-case radio interferometry scenario of near future, detecting the effects of DM in our galaxy with imaging of Sgr A$^\star$ will be nearly impossible. 

Although this limited analysis suggests that detection of the effects of the DM spike with BH imaging in the near future is unlikely, it could become detectable in the far future. As the infinite-resolution images showed, the imprint of the DM spike in both strong and unique, and with advanced detectors (e.g., strategically-enhanced ground VLBI~\citep{Blackburn:2019bly}, space VLBI~\citep{Fromm:2021flr}) and detection techniques (e.g., variational image feature extraction~\citep{Tiede:2020iif}), the DM effects may become resolvable.

\begin{deluxetable}{lccc}[t]
\tablecaption{A comparison of SSIM values for different pairs of reconstructed images. The images are denoted by their $(r_{\textrm{in}}, a_*, \rho)$ values. For the top two rows, image $B$: $(r_{\textrm{ISCO}}, 0.99, 0)$ and image $C$: $(r_{\textrm{ISCO}}, 0, 10^5\rho_{sp})$, whereas for the bottom two rows, image $B$: $(r_{\textrm{ph}}, 0.99, 0)$ and image $C$: $(r_{\textrm{ph}}, 0, 10^5\rho_{sp})$.}
\tablehead{\colhead{{Image $A$}} & \colhead{{SSIM($B,A$)}} & \colhead{{SSIM($C,A$)}} & \colhead{{SSIM($C,B$)}}}
\startdata
$r_{\textrm{ISCO}},\; 0,\; 0$ & $0.998$ & $0.999$ & $0.998$ \\
$r_{\textrm{ISCO}},\; 0.99,\; 10^5\rho_{sp}$ & $0.999$ & $0.998$ & $0.998$ \\
$r_{\textrm{ph}},\; 0,\; 0$  & $0.999$ & $0.999$ & $0.999$  \\
$r_{\textrm{ph}},\; 0.99,\; 10^5\rho_{sp}$  & $0.999$ & $0.999$ & $0.999$\\
\enddata
\end{deluxetable}
\label{tab:ssim-rec}


\section{\label{sec:conclude}Conclusions}

In this work, we have explored the effects of a dark matter spike on various observables relevant for the Sgr A$^\star$ BH, and identified whether current and/or future astrophysical observations of Sgr A$^\star$ could detect the presence of such DM spike. Using a density model for the DM spike, we have first constructed the spacetime metric around a static and spherically symmetric BH and then generalized it to the rotating case. 

For the special case of a static BH immersed in DM spike, we determined and analyzed the constraints using the S2 star orbit features around the Sgr A$^\star$ BH, on the two free parameters of the spacetime which characterize the density $\rho_\text{b}$ and the innermost boundary $r_\text{b}$ of the DM distribution. For the two specific values of $\gamma_\text{sp}$, we have shown that $r_\text{b}\sim 9$ in the BH mass units, while near the BH $\rho_\text{b}\sim 10^{-7}$ in the cgs units. These constraints indicate that DM in our galaxy can be close to the central BH with a relatively high density.

We calculate the effect of the presence of the DM on the BH shadow. We show that by increasing the DM spike density the shadow radius increases and, for very high DM densities, can grow significantly large.  Within a highly idealized scenario of a geometrically-thick and optically-thin accretion disk model around the Sgr A$^\star$ BH, which is radiating as a monochromatic power-law, we also analyzed BH images. Our results show that for the available observational data for the DM spike density $\rho_\text{sp} \sim 10^{-23}-10^{-24}$ g/cm$^3$ the effect of the DM is small, however, when the DM density is of the order $\rho_\text{sp} \sim  (10^{-19}-10^{-20})$ g/cm$^3$, the image is strongly modified. The modification introduced by the DM is, moreover, unique and can be distinguished from the effects of BH spin. 

To probe detectability of the DM effects, we simulate observations of a Sgr A$^\star$-DM spike system with an EHT-like array potentially realizable in the near future. Even for optimistic scenarios, we find that the effects of the DM are unlikely to be detectable in the near future. This may change in the far future with better detectors and detection techniques.

\acknowledgements
MJ would like to thank Cosimo Bambi and Jorge Rueda for helpful discussions during the preparation of this work. The work of QW and MJ is supported in part by the National Key Research and Development Program of China Grant No.2020YFC2201503, the Zhejiang Provincial Natural Science Foundation of China under Grant No. LR21A050001, the Zhejiang Provincial Natural Science Foundation of China under Grant No.LY20A050002, the Fundamental Research Funds for the Provincial Universities of Zhejiang in China under Grant No. RF-A2019015, and National Natural Science Foundation of China under Grant No. 11675143. SN acknowledges support from the Alexander von Humboldt Foundation.

\bibliographystyle{aasjournal}
\bibliography{references}


\end{document}